\documentclass[12pt,a4paper]{article}

\usepackage{amsmath}
\usepackage{amssymb}
\usepackage{float}
\usepackage{tabularx}
\usepackage{times}
\usepackage{theorem}
\usepackage{graphicx}
\usepackage{graphics}
\usepackage[usenames]{color}
\usepackage{hyperref}

\theorembodyfont{\rmfamily \itshape} \newtheorem{conj}{Conjecture}[section]
\theorembodyfont{\rmfamily \itshape} 
\theorembodyfont{\rmfamily \itshape} \newtheorem{defn}[conj]{Definition}
\theorembodyfont{\rmfamily \itshape} \newtheorem{lemma}[conj]{Lemma}
\theorembodyfont{\rmfamily \itshape} \newtheorem{prop}[conj]{Proposition}
\theorembodyfont{\rmfamily \itshape} \newtheorem{thrm}[conj]{Theorem}
\theoremheaderfont{\bf\scshape}

\newcommand{\Arctan}{\operatorname{Arctan}}
\newcommand{\Arg}{\operatorname{Arg}}

\newcommand{\tr}{\operatorname{tr}}
\newcommand{\refeq}[1]{(\ref{#1})}
\newcommand{\reffig}[1]{Figure~\ref{#1}}
\newcommand{\reftab}[1]{Table~\ref{#1}}

\newcommand{\refsec}[1]{Section~\ref{#1}}
\newcommand{\lpar}{\parallel \negthickspace}
\newcommand{\rpar}{\negthickspace \parallel}

\newcommand{\proofend}{\hfill$\Box$\newline}

\title{Pure point spectrum for the time-evolution of a periodically rank-N
kicked Hamiltonian}
\author{
  James~McCaw\thanks{Electronic mail: j.mccaw@physics.unimelb.edu.au}
  \phantom{a}and B.~H.~J.~McKellar\thanks{Electronic mail: b.mckellar@physics.unimelb.edu.au} \\
  \textit{School of Physics, Research Centre for High Energy Physics,} \\
  \textit{The University of Melbourne, Victoria, 3010, Australia.}
}

\date{(Dated: \today)}

\begin{document}

\maketitle


\begin{abstract}
We find the conditions under which the spectrum of the unitary time-evolution
operator for a periodically rank-N kicked system remains pure point. This
stability result allows one to analyse the onset of, or lack of chaos in
this class of quantum mechanical systems, extending the results for
rank-$1$
systems produced by Combescure and others. This work includes a number of
unitary theorems equivalent to those well known and used in the self-adjoint
theory.
\end{abstract}


\section{\label{sec:r_int}Introduction}

We will derive
conditions on the time-periodic perturbations to the base Hamiltonian for the
spectrum of the Floquet operator to remain pure point.
We consider Hamiltonians of the form
\begin{equation}
\label{eq:r_hamiltonian}
  H(t) = H_0 + A^*WA \sum_{n=0}^\infty \delta(t-nT)
\end{equation}
where $A$ is bounded, $W$ is self adjoint and $H_0$ has pure point
(discrete) spectrum. The time-evolution of such Hamiltonians is of great
interest in quantum chaos, and of central importance is the spectral
properties of the Floquet operator, defined as
\begin{equation}
\label{eq:r_floqop}
V=e^{iA^*WA / \hbar}e^{-iH_0 T / \hbar}
\end{equation}
which comes directly from considering the time-evolution of the kicked
system
\begin{equation*}
U(t) = \left[
\exp\left(-\frac{i}{\hbar}\int_0^t dt'H(t') \right)
\right]_+
\end{equation*}
with $H(t')$ given by \refeq{eq:r_hamiltonian}. The spectrum of the Floquet
operator is known as the ``quasi-energy spectrum''.

This work is an extension of a result of Combescure \cite{Combescure90}.
Our results are based on the self-adjoint work by Howland \cite{Howland87}. 
If we choose $A$ to be a rank-$1$ perturbation,
\begin{equation*}
\begin{split}
A &= | \psi\rangle\langle\psi | \\
W &= \lambda I
\end{split}
\end{equation*}
we reproduce the work of Combescure \cite{Combescure90}. The vector
$|\psi\rangle$ is a linear combination of orthonormal basis states,
$|\phi_n\rangle$ of the unperturbed Hamiltonian $H_0$
\begin{equation}
|\psi\rangle = \sum_{n=0}^\infty a_n|\phi_n\rangle\text{.}
\end{equation}
Combescure showed that if $\psi \in l_1(H_0)$, that is if
\begin{equation}
\sum_{n=0}^\infty |a_n| < \infty
\end{equation}
then the quasi-energy spectrum remains pure point for almost every
perturbation strength $\lambda$. We will generalise this result to all
finite rank perturbations
\begin{equation}
\begin{split}
\label{eq:r_rankNpert}
A &= \sum_{k=1}^N A_k = \sum_{k=1}^N | \psi_k\rangle\langle\psi_k | \\
W &= \sum_{k=1}^N \lambda_k | \psi_k\rangle\langle\psi_k |
\end{split}
\end{equation}
where $\lambda_k \in \mathbb{R}$ and each vector $|\psi_k\rangle$ is a
linear combination of the $H_0$ basis states, $|\phi_n\rangle$
\begin{equation}
\label{eq:r_psi_k}
|\psi_k\rangle = \sum_{n=0}^\infty (a_k)_n | \phi_n\rangle\text{.}
\end{equation}
The states $| \psi_k\rangle$ are orthogonal
\begin{equation}
\label{eq:r_orth_states}
\langle\psi_k | \psi_l\rangle = \delta_{kl}\text{.}
\end{equation}

The basic result is that if each $|\psi_k\rangle$ is in $l_1(H_0)$, the
spectrum of $V$ will remain pure point for almost every perturbation
strength.

The perturbation for which we prove that the quasi-energy spectrum remains
pure point is in fact more general than the finite rank perturbation
presented above. The finite rank result is however the motivation for
undertaking this work.

Howland \cite{Howland87} showed that the Hamiltonian \refeq{eq:r_hamiltonian}
has a pure point spectrum if the $\psi_k$s are in $l_1(H_0)$. Here, we
follow a similar argument, showing that the continuous part of the spectrum
of $V$ is empty, allowing us to conclude that the spectrum of $V$ must be
pure point.

Associated with what we have termed the Floquet Operator, is the ``Floquet
Hamiltonian''
\begin{equation*}
K = -id/dt + H(t)\text{.}
\end{equation*}
It turns out that $K$ provides a different way to access similar
information to what we are seeking. Developed in papers
by Howland \cite{Howland74}, \cite{Howland79} and \cite{Howland89.1} and
linked to our Floquet operator in \cite{Bunimovich} (p.~808), $K$ was
introduced, in some part, because directly working with $V$ proved too
difficult. The large body of knowledge on self-adjoint operators provides a
mature basis for proving theorems about $K$. As discussed in
\cite{Bunimovich}, the spectrum of $K$ is easily related algebraically to
that of $V$, so results on the spectrum for $K$ and $V$ are equivalent.

Working directly with $V$, however, as we do here, is valuable in that
it gives a transparent, direct insight into the dynamics of the perturbed
system $H(t)$. After the completion of this work, which is a unitary
equivalent to that of Howland \cite{Howland87}, we discovered that Howland had
used his work \cite{Howland87} on the spectrum of self-adjoint operators to
obtain similar results to what we do here \cite{Howland89.1}.

The relationship between our work and Howland's works \cite{Howland87} and
\cite{Howland89.1} is similar to the relationship between the self-adjoint
rank-$1$
work of Simon and Wolff \cite{Simon86} and the unitary rank-$1$ work of
Combescure \cite{Combescure90}.

The techniques developed in this paper provide new, general theorems
applicable to unitary operators and show that it is possible to develop
the theory of the spectrum of time-evolution operators directly, without
need for the techniques of \cite{Howland74} briefly mentioned here.


\subsection{\label{ssec:r_motivation}Motivation}

The classical study of chaos is now a well established and flourishing
field of research in mathematics and mathematical physics. Chaotic behaviour
seems to pervade a vast spectrum of dynamical systems, and an appreciation
of it is essential for a detailed understanding of such systems. The
classic example of the earth's weather patterns always comes to mind when
chaos is mentioned.

The microscopic world, however, is not governed by the laws of classical
dynamics. In the realm of small quantum numbers the dynamics of a system
is governed by the Schr\"odinger
equation. In such systems, the simple and elegant definitions of chaos
such as positive Lyapunov exponent, which hold for classical systems,
are not applicable. In fact, there
is no universally accepted definition of quantum chaos. Some model systems
show what many would consider ``chaotic behaviour'', yet there are general
arguments made by some \cite{Partovi} to the effect that ``quantum chaos''
does not exist. Our study of one aspect of ``quantum chaos'' is motivated by
much of this work. Taking note of these uncertainties and conflicting views,
two questions arise that are of central importance.
\begin{enumerate}
\item What properties of a quantum mechanical system determine whether or
not the corresponding classical system that derives from it will display
chaotic behaviour?
\item Are there in fact quantum systems that display chaotic behaviour at
the quantum level?
\end{enumerate}

The former question is intimately linked to the ``Correspondence
Principle'' and theories of quantum measurement. Needless to say, this area
of fundamental physics is infamous for its interpretational difficulties
and seemingly inconsistent behaviour.

The latter question too, is the source of much debate in the literature.
As in any immature study, quantum chaos is struggling to be self
consistently defined. A wide range of possible definitions and
interpretations of what quantum chaos actually is have been put forward,
many in direct contradiction with one another. At some stage in the future
presumably, we will find a satisfactory criteria for what constitutes
quantum chaos. Until then, many attempts to look at particular aspects of
the dynamics of quantum systems will be (and have been) made. Some papers,
courtesy of their definition of quantum chaos, come to the conclusion that
there is no such thing as quantum chaos. That is, they conclude that no
quantum system can display chaotic behaviour. Other papers, simply as a
consequence of a different starting point, come to the conclusion that
there are quantum systems that display chaos.

There are general arguments that allow one to categorise the behaviour of a
quantum system based upon the spectral composition of the quasi-energy
spectrum. Hence, Combescure's work on the spectrum is relevant to the study
of chaos. Our work, by extending the result of Combescure, may allow for the
further categorisation of classes of Hamiltonian systems as chaotic or
otherwise. This is further discussed in the next section.


\subsection{\label{ssec:r_chaos}Spectral analysis of operators and a link
to chaos}

The intuitive definition for the energy spectrum of a quantum system is
best seen through example, say the hydrogen atom. The bound states of
hydrogen are a countable number of isolated, discrete energies. Each energy
corresponds to an eigenvalue of the system and the set of these points
makes up the point energy spectrum. The positive energy scattering states
form the continuous energy spectrum. Thus, the energy spectrum for the
hydrogen system consists of two disjoint parts: the negative energy
discrete (or ``point'') spectrum, and the positive energy continuous
spectrum.

For hydrogen, $\sigma_p(H) = \{\alpha_n ; \alpha_n \approx -13.6/n^2
\text{ for } n\in \mathbb{N}\}$, and $\sigma_{cont}(H) = (0,\infty)$.

As another simple example, the harmonic oscillator quantum system has only
discrete energy levels, and thus is said to be ``pure point''. That is, the
eigenvectors of the harmonic oscillator form a basis of the Hilbert Space.

While these simple examples have shown clearly that we can split the energy
spectrum into point and continuous parts, this is not the whole story. The
mathematical treatment of operators and measures shows that the spectrum in
fact consists of three parts, the point, absolutely continuous and
singularly continuous spectrum. For an appreciation of the work that
follows, a mathematically rigorous understanding of the spectra is
necessary. The introductory chapters in \cite{Reed1} are essential reading.

One must note that the concept of spectrum is associated with a particular
operator.  Typically, we talk of the energy spectrum, associated with the
Hamiltonian.  However, all operators (e.g.,\ Hamiltonian, Floquet etc.) have
a spectrum. A failure to realise this has lead to a number of confused
papers (see for example \cite{Milek90.1}) which use results on the spectrum
of the Hamiltonian in a discussion of the spectrum of the Floquet operator.
With these words of warning, we return to a discussion of the Floquet (or
quasi-energy) spectrum of a quantum system.

The link between spectral properties and dynamics is an active field of
research and is not yet fully understood. The introduction to the paper of
Y.~Last \cite{Last} provides an informative overview of the field and gives
details on some of the most relevant theorems and results, including the
RAGE theorem (\cite{Reed3}, p.~341, Theorem~XI.115). See also
\cite{Combes}. Y.~Last's paper deals with systems where the Hamiltonian
spectrum is of interest. In time-periodic Hamiltonian systems, the spectrum
of the Floquet operator takes over that role. K.~Yajima and H.~Kitada
\cite{Yajima} show that RAGE-like results apply to time-periodic systems,
as we have here, and thus an analysis of the Floquet operator spectrum
is of interest.

Refering to either the Hamiltonian spectrum or the Floquet spectrum where
appropriate, and the appropriate RAGE-like theorem, we now comment on the
``typical'' manifestation of the spectrum. A typical quantum mechanical
system does not posses a singularly continuous spectral component and thus,
singular continuity is not usually mentioned in texts on quantum mechanics.
This however, is not to say that it can't exist, or that it doesn't
manifest itself in the dynamical behaviour of appropriate systems. With an
understanding that Milek and Seba meant to refer to the RAGE-like theorem
in \cite{Yajima} rather than the RAGE theorem itself, the argument
presented in Section~II of their paper \cite{Milek90.1} shows that if a
system possesses a singularly continuous quasi-energy spectrum then its
energy growth over time \emph{may} be characteristic of a classically
chaotic system. Thus, establishing the existence or otherwise of singular
continuous spectra for the Floquet operator can be seen as of central
importance to the question of whether or not a quantum mechanical system is
chaotic. It must be noted that the arguments presented by Milek and Seba
are acknowledged to be anything but rigorous---a point clearly established
by Antoniou and Suchanecki \cite{Antoniou02,Antoniou03}.

It is with the application of the RAGE-like theorem in mind \cite{Yajima},
that we undertook the following work on the analysis of the quasi-energy
spectrum of the class of Hamiltonians as defined by
\refeq{eq:r_hamiltonian}. The aforementioned work by Milek and Seba
\cite{Milek90.1}, utilising the rank-$1$ work of Combescure, has shown the
manifestation of singularly continuous spectra in numerical simulations of
rank-$1$ kicked rotor quantum systems. The work here has the potential to
extend upon this, and provide a rigorous mathematical basis to numerical
calculations on the time-evolution of higher rank kicked quantum systems.


\subsection{\label{ssec:r_outline}Outline and summary of results}

In \refsec{sec:r_spectra} we will present the main theorems of the
paper, concerned with establishing when systems of the form given by
\refeq{eq:r_hamiltonian} maintain a pure point quasi-energy
spectrum. Parallelling Howland's paper \cite{Howland87} on self-adjoint
perturbations of pure point Hamiltonians, the key ideas are those of
$U$-finiteness and the absolute continuity of the multiplication
operator $\mathbb{V}$. To establish the second of these concepts for our
unitary case (remember that we are concerned with the spectral properties
of the unitary time-evolution operator and not with the spectral properties
of the self adjoint Hamiltonian itself), we require a modified version of
the Putnam--Kato theorem \cite{Reed4}. This, and associated theorems are the
topic of \refsec{sec:r_pk}. \refsec{sec:r_finite} uses the results of
\refsec{sec:r_spectra} and \refsec{sec:r_pk} to give the final results,
which are then discussed in \refsec{sec:r_discussion}.


\subsection{\label{ssec:r_notation}Notation}

We inherit our notation directly from the work of Howland \cite{Howland87}.
$\mathcal{H}$ and $\mathcal{K}$ will denote Hilbert spaces throughout this
paper. They will always be separable. The inner product of two vectors $x$
and $y$ is $\langle x,y \rangle$, and the norm of a vector $x$ is
$\lpar x \rpar \;=\langle x,x \rangle^{1/2}$.
For an operator $A:\mathcal{H}\rightarrow\mathcal{K}$ we define
\begin{itemize}
\item the domain $D(A)$ ; the vectors $x\in\mathcal{H}$ for which $Ax$ is
defined,
\item the range $R(A) = \left\{y\in\mathcal{K} : y=Ax \text{ for some
$x\in\mathcal{H}$} \right\}$,
\item the kernel ker $A = \left\{ x\in\mathcal{H} : Ax = 0 \right\}$, and
\item the operator norm $\lpar A \rpar \;=
  \sup_{x\in D(A)\;:\; \parallel x\parallel=1}\left\{\lpar Ax \rpar \right\}$.
\end{itemize}
For any set $S\in\mathbb{C}$, $\overline{S}$ is the closure of $S$. If
$A_n$ is a sequence of operators, s-lim~$A_n$ (also
$A_n\overset{s}{\rightarrow}A$) denotes the strong limit,
$\lpar (A_n - A)g \rpar \rightarrow 0$ for all $g\in\mathcal{H}$.
w-lim~$A_n$ (also $A_n\overset{w}{\rightarrow}A$) denotes the weak limit,
$| \langle A_ng,f \rangle - \langle Ag,f \rangle | \rightarrow 0$ for all
$g,f \in\mathcal{H}$. By the Schwartz inequality, the weak limit exists if
the condition above is satisfied for $f=g$. We will also have need for the
norm limit of an operator, $\lpar A_n - A \rpar \rightarrow 0$.
  
For a unitary operator
$V= \int e^{-i\theta} E(d\theta)$ on $\mathcal{H}$, we define for any Borel
set $S$, $E[S] = \int_S E(d\theta)$. The $E(d\theta)$ are orthogonal
projection operators, i.e.,\ $E^2 = E$ and thus
\begin{equation*}
\int \left|f(\theta)\right|^2E(d\theta)
 = \left|\int f(\theta)E(d\theta)\right|^2\text{.}
\end{equation*}
We decompose our operator into its pure point ($V^p$), singular continuous
($V^{sc}$) and absolutely continuous ($V^{ac}$) components.
$V^s = V^p + V^{sc}$ is the singular part of the operator $V$. Similarly, we
define the corresponding spectral measures $E^p$, $E^{sc}$, $E^{ac}$ and
$E^s$. For a vector $x\in\mathcal{H}$, $m_x$ is the measure
\begin{equation*}
m_x(S) = \langle E(S)x,x \rangle\text{.}
\end{equation*}
Again, we define $m_x^p$, $m_x^{sc}$, $m_x^{ac}$ and $m_x^s$. See
(\cite{Reed1}, p.~19--23) for an excellent description of these.  By their
definition, $m_x^p$, $m_x^{sc}$ and $m_x^{ac}$ are mutually singular, so we
may write the Hilbert space as a direct sum (i.e.,\ each of the spaces below
is invariant)
\begin{equation*}
\mathcal{H} 
  = \mathcal{H}_{pp} \oplus \mathcal{H}_{ac} \oplus
    \mathcal{H}_{sc}\text{.}
\end{equation*}
The spectrum
of $V$ is $\sigma(V)$, defined by
\begin{equation*}
\sigma(V) = \{ \alpha \in \mathbb{C} : \alpha I - V \text{ is not
invertible}\}\text{.}
\end{equation*}
If $Tx=\alpha x$ for some $x\in\mathcal{H}$ and $\alpha \in \mathbb{C}$,
then $x$ is an eigenvector, with corresponding eigenvalue $\alpha$.
The closure of the set of $\alpha$s forms the point spectrum of $V$
\begin{equation*}
\begin{split}
\sigma_p(V)
  &= \overline{\{ \alpha : \alpha \text{ an eigenvalue of $V$}\}} \\
  &= \sigma(V \upharpoonright \mathcal{H}_p)\text{.}
\end{split}
\end{equation*}
We say that $V$ is \emph{pure point} if and only if
the eigenvectors of $V$ form a basis of $\mathcal{H}$.
The absolutely (singularly) continuous spectrum, $\sigma_{ac(sc)}(V)$ is
similarly defined by
\begin{equation*}
\sigma_{ac(sc)}(V) = \sigma(V \upharpoonright \mathcal{H}_{ac(sc)})\text{.}
\end{equation*}
For a complete discussion and analysis of these topics, the most convenient
reference is (\cite{Reed1}, p.~19--23, 188, 230--231) or \cite{Kato76}.

A set $\mathcal{S}\in\mathcal{H}$ is said to reduce an operator $A$ if both
$\mathcal{S}$ and its ortho-complement $\mathcal{H} \ominus \mathcal{S}$
are invariant subspaces for $A$.

A vector $\phi$ is cyclic for an operator $A$ if and only if finite linear
combinations of elements of $\{A^n\phi\}_{n=0}^\infty$ are dense in
$\mathcal{H}$. This motivates the definition that a set $\mathcal{S}$ is
cyclic for $\mathcal{H}$ if and only if the smallest closed reducing
subspace of $\mathcal{H}$ containing $\mathcal{S}$ is $\mathcal{H}$.

We will also use some basic set notation. $A \cap B$ and $A \cup
B$ are, as usual, the intersection and union of sets $A$ and $B$
respectively. $A^\text{c}$ is the complement of $A$. $A \sim B$ is
$A \cap B^\text{c}$.
Note that $(A \sim B) \sim C$ is not equal to $A \sim (B \sim C)$, the
former being a subset of the latter.

The function $\chi_{\text{\tiny\itshape S}}(x)$ is the characteristic
function for a set $S$.


\section{\label{sec:r_spectra}Spectral properties of the Floquet operator}

Let $U$ be unitary on $\mathcal{H}$ and let $\mathcal{K}$ be an auxiliary
Hilbert space. Define the closed operator
$A:\mathcal{H}\rightarrow\mathcal{K}$, with dense domain $D(A)$. For our
purposes, $A$ bounded on $\mathcal{H}$ is adequate. We work with a
modification (multiplication by $e^{i\theta}$) of the resolvent of $U$
\begin{equation}
\label{eq:r_Fdef}
F(\theta;U) = \left(1-Ue^{i\theta}\right)^{-1}
\end{equation}
and define for $\theta \in [0,2\pi)$, and $\epsilon > 0$ the
function $G_\epsilon:\mathcal{K}\rightarrow\mathcal{K}$
\begin{equation}
\label{eq:r_Geps}
G_\epsilon(\theta;U,A)
  = AF^*(\theta_+;U)F(\theta_+;U)A^*\text{.}
\end{equation}
where $\theta_\pm = \theta \pm i \epsilon$. Let $J$ be a subset of $[0,2\pi)$. 


\begin{defn}[U-finite]
\label{defn:1}
The operator $A$ is $U$-finite if and only if the operator
$G_\epsilon(\theta;U,A)$ has a bounded extension to $\mathcal{K}$, and
\begin{equation}
\label{eq:r_strongG}
G(\theta;U,A) = \text{\textup{s-}}\underset{\epsilon\downarrow 0}{\lim}
  \;G_\epsilon(\theta;U,A)
\end{equation}
exists for a.e.\ $\theta \in J$.
\end{defn}

We define the function
\begin{equation}
\begin{split}
\label{eq:r_delta_epsilon}
\delta_\epsilon(t) &= \frac{1}{2\pi}\left(
    \sum_{n=0}^{\infty}e^{in(t+i\epsilon)} +
    \sum_{n=-\infty}^0 e^{in(t-i\epsilon)} - 1\right) \\
  &= \frac{1}{2\pi}\frac{1-e^{-2\epsilon}}
    {1-2e^{-\epsilon}\cos(t)+e^{-2\epsilon}}\text{.}
\end{split}
\end{equation}
The limit as $\epsilon\rightarrow 0$ of $\delta_\epsilon(t)$ is a series
representation of the $\delta$-function. The proof is based on showing that
\begin{equation*}
\underset{\epsilon\downarrow 0}{\lim}
  \int_{-\pi}^\pi g(t)\delta_\epsilon(t)dt = 0
\end{equation*}
where $g(t) = f(t) - f(0)$ and $f(t)$ is bounded in $(-\pi,\pi)$. One splits
the integral into three parts,
$\int_{-\pi}^{-\xi} + \int_{-\xi}^\xi + \int_\xi^\pi$. One must assume that
$f(t)$ is continuous at $t=0$ (otherwise $\int f(t)\delta(t)dt$ is not well
defined) so that
\begin{equation*}
\forall\eta, \exists \xi > 0 \text{ s.t. }
\forall t, |t|<\xi \text{ we have } |f(t)-f(0)|<\eta\text{.}
\end{equation*}
The assumption that $f(t)$ is bounded on $(-\pi,\pi)$ is also required.

The first and third integrals are zero because
$\delta_\epsilon(t) \rightarrow 0$ for $t\neq 0$
from \refeq{eq:r_delta_epsilon} and the assumption that $g(t)$ is bounded. The
second integral from $-\xi$ to $\xi$ is zero by the continuity of $f(t)$ at
$t=0$, the positivity of $\delta_\epsilon(t)$ and (\cite{Gradshteyn}, p.~435,
(3.792.1)).

Given \refeq{eq:r_delta_epsilon} and the spectral decomposition of $U$, we
may write
\begin{equation}
\begin{split}
\label{eq:r_delta_resolvent}
\delta_\epsilon\left(1-Ue^{i\theta}\right)
  &= \int\delta_\epsilon\left(1-e^{i(\theta-\theta')}\right)E(d\theta') \\
  &= \int\delta_\epsilon(\theta-\theta')E(d\theta') \\
  &= \frac{1}{2\pi}\left(1-e^{-2\epsilon}\right)
    F^*(\theta_+;U)F(\theta_+;U)\text{.}
\end{split}
\end{equation}

The existence of a non-trivial $U$-finite operator will have important
consequences for the spectrum of the Floquet operator $V$. We introduce the
set
\begin{equation*}
N(U,A,J) = \{\theta\in J : 
  \text{s-}\underset{\epsilon\downarrow 0}{\lim}
  G_\epsilon(\theta;U,A) \text{ does not exist} \}
\end{equation*}
of measure zero, which enters the theorem. We will often refer to
this set simply as $N$ during proofs.


\begin{thrm}
\label{thrm:r_2}
If $A$ is $U$-finite on $J$ and $R(A^*)$ is cyclic for $U$, then
\begin{enumerate}
\item $U$ has no absolutely continuous spectrum in $J$, and
  \label{thrm:r_2a}
\item the singular spectrum of $U$ in $J$ is supported by $N(U,A,J)$.
  \label{thrm:r_2b}
\end{enumerate}
\end{thrm}


\emph{Proof.} (\ref{thrm:r_2a}) Following Howland, we note that the
absolutely continuous spectral measure, $m_y^{ac}(J)$ is the
$\epsilon\rightarrow 0$ limit of
$\left\langle\delta_\epsilon\left(1-Ue^{i\theta}\right)y,y\right\rangle$
for $\theta \in J$.
If $y\in\mathcal{H}$ is in $R(A^*)$, allowing us to write $y=A^*x$ for some
$x\in\mathcal{K}$, then
\begin{equation*}
\begin{split}
\underset{\epsilon\downarrow 0}{\lim}
  &\left\langle\delta_\epsilon\left(1-Ue^{i\theta}\right)y,y\right\rangle \\
  &= \underset{\epsilon\downarrow 0}{\lim}
    \left\langle\delta_\epsilon\left(1-Ue^{i\theta}\right)A^*x,A^*x
    \right\rangle \\
  &= \underset{\epsilon\downarrow 0}{\lim}\frac{\epsilon(1-\epsilon)}{\pi}
    \left\langle G_\epsilon(\theta;U,A)x,x\right\rangle = 0
\end{split}
\end{equation*}
for a.e.\ $\theta\in J$. The set $\mathcal{Y}$ of vectors $y$ for which
$m_y^{ac}(J)=0$ is a closed reducing subspace of $\mathcal{H}$, and by
construction contains the cyclic set $R(A^*)$ as a subset. Because
$\mathcal{Y}$ is invariant, finite linear combinations of action with $U^n$
leaves us in $\mathcal{Y}$. Due to the cyclicity, these same linear
combinations allow us to reach any $y\in\mathcal{H}$. Thus, the set
$\mathcal{Y}$ of vectors $y$ with $m_y^{ac}(J)=0$ must be the whole Hilbert
space $\mathcal{H}$. So there is no absolutely continuous spectrum of $U$
in $J$.

\vspace{\baselineskip}


(\ref{thrm:r_2b}) A theorem of de~la~Vall\'{e}e Pousin (\cite{Saks}, p.~127,
(9.6)) states that the singular part of the spectrum of a function is
supported on the set where the derivative is infinite. In our case, this
corresponds to finding where $m_y(d\theta)\rightarrow\infty$. We calculate
\begin{equation*}
\begin{split}
\underset{\epsilon\downarrow 0}{\lim}
  \langle \delta_\epsilon\left(1-Ue^{i\theta}\right)y,y\rangle
  &= \int\delta(\theta - \theta') \langle E(d\theta')y,y\rangle \\
  &= \int\delta(\theta - \theta') m_y(d\theta') \\
  &= m_y(d\theta)\text{.}
\end{split}
\end{equation*}

Thus, $m_y^s = m_y^{sc} + m_y^{pp}$ is supported on the set where
\begin{equation}
\label{eq:r_vallee}
\underset{\epsilon\downarrow 0}{\lim}
  \left\langle\delta_\epsilon\left(1-Ue^{i\theta}\right)y,y\right\rangle
  = \infty\text{.}
\end{equation}

From the proof to part (\ref{thrm:r_2a}), if $y=A^*x$ then the limit
\refeq{eq:r_vallee} is zero for $\theta \in J$, $\theta \notin N$, so $m_y^s$
in $J$ must be supported by $N$. The set of vectors $y$ with
$m_y^s(J \cap N^c) \equiv m_y^s (J \sim N) = 0$ is closed, invariant and
contains $R(A^*)$, so must be $\mathcal{H}$ by the argument above. Thus,
the singular spectrum of $U$ is supported on the set $N$.\proofend


We now define a new operator, $Q(z):\mathcal{K}\rightarrow\mathcal{K}$
\begin{equation*}
Q(z) = A(1-Uz)^{-1}A^*\text{.}
\end{equation*}

Note that
\begin{equation}
\label{eq:r_QfromF}
Q\left(e^{i\theta_\pm}\right) = AF(\theta_\pm;U)A^*\text{.}
\end{equation}
$Q(z)$ is clearly well defined for $|z|\neq 1$. Proposition~\ref{prop:r_3}
shows that the definition can be extended to $|z|=1$.


\begin{prop}
\label{prop:r_3}
Let $A$ be bounded. If $\theta \in J$, but $\theta \notin N(U,A,J)$, then
\begin{enumerate}
\item the operator $Q\left(e^{i\theta}\right)
  = A\left(1-Ue^{i\theta}\right)^{-1}A^*$
  is bounded on $\mathcal{K}$, and
  \label{prop:r_3a}
\item one has $\text{\textup{s-}}\underset{\epsilon\downarrow 0}{\lim}
  Q\left(e^{\pm i(\theta\pm i\epsilon)}\right)
  = Q\left(e^{\pm i\theta}\right)$.
  \label{prop:r_3b}
\end{enumerate}
\end{prop}


\emph{Proof.} (\ref{prop:r_3a}) Without loss of generality, take
$\theta = 0$ ($z=1$). By Theorem~\ref{thrm:r_2},
$e^{-i0} \notin \sigma_p(U)$, so $(1-Ue^{i0})^{-1}$ exists as a
densely defined operator. As $A$ is a bounded operator, it suffices to show
that $\left(1-Ue^{i0}\right)^{-1}A^*$ is bounded. We have
\begin{equation}
\begin{split}
\label{eq:r_Qpbounded}
  \lpar \left(1-Ue^{i0_+}\right)^{-1}A^*x \rpar^2
  &= \langle F(0_+;U)A^*x, F(0_+;U)A^*x \rangle \\
  &= \langle AF^*(0_+;U)F(0_+;U)A^*x,x \rangle \\
  &= \langle G_\epsilon(0;U,A)x,x \rangle \leq C|x|^2
    \text{ (as $\theta \notin N$)}
\end{split}
\end{equation}
for some real constant $C$.
If $y=A^*x$, noting $U = \int e^{-i\theta}E(d\theta)$, we also have
\begin{equation*}
  \lpar \left(1-Ue^{i0_+}\right)^{-1}A^*x \rpar^2 =
    \int\left(\frac{1}{1-e^{-i\theta}e^{-\epsilon}}\right)
    \left(\frac{1}{1-e^{i\theta}e^{-\epsilon}}\right)
    \langle E(d\theta)y,y\rangle\text{.}
\end{equation*}
In light of \refeq{eq:r_Qpbounded}, we may safely take $\epsilon$ to zero,
to obtain
\begin{equation}
\label{eq:r_Q_bounded}
\int\left(\frac{1}{1-e^{-i\theta}}\right)\left(\frac{1}{1-e^{i\theta}}\right)
  \langle E(d\theta)y,y \rangle \leq C\lpar x \rpar^2 < \infty\text{.}
\end{equation}
From \refeq{eq:r_Q_bounded}, we have
\begin{multline}
\int\left(\frac{1}{1-e^{-i\theta}}\right)\left(\frac{1}{1-e^{i\theta}}\right)
  \langle E(d\theta)y,y \rangle \\
  = \left\langle [1-U]^{-1}y,[1-U]^{-1}y\right\rangle \leq C\lpar x \rpar^2
    < \infty
\end{multline}
so $y\in D\left[(1-U)^{-1}\right]$. Thus, $Q(1) = A(1-U)^{-1}A^*$
is defined on all $\mathcal{K}$ and bounded.
\vspace{\baselineskip} 


(\ref{prop:r_3b}) For $y \in D\left((1-U)^{-1}\right)$, we show that the
difference between $Q\left(e^{\pm i(0 \pm i\epsilon)}\right)$ and
$Q\left(e^{\pm i0}\right)$ tends to zero as $\epsilon\rightarrow 0$. Again,
due to the boundedness of $A$, we need only show that
\begin{equation*}
\lpar \left( \left(1-Ue^{i0_+}\right)^{-1} - \left(1-U\right)^{-1} \right)
  A^*x \rpar
\end{equation*}
tends to zero. Consider
\begin{equation}
\begin{split}
\label{eq:r_lim_Q_exists}
\bigl| &\bigl(1-Ue^{-\epsilon}\bigr)^{-1}y - (1-U)^{-1}y \bigr|^2 \\
  &=\int \left|\frac{1}{1-e^{-i\theta}e^{-\epsilon}}-
    \frac{1}{1-e^{-i\theta}} \right|^2 \langle E(d\theta)y,y \rangle \\
  &= \int \left(
    \frac{\left(1-e^{-\epsilon}\right)^2}
    {1-2e^{-\epsilon}\cos\theta+e^{-2\epsilon}}\right)
    \frac{\langle E(d\theta)y,y \rangle}
    {\left(1-e^{-i\theta}\right)
    \left(1-e^{i\theta}\right)}\text{.}
\end{split}
\end{equation}

The first factor is bounded and tends to zero for $\theta \neq 0$.
The second factor is the measure from \refeq{eq:r_Q_bounded}. Clearly, away
from the origin, the integral tends to zero. About
the origin, we must take some care to show that there is no contribution to
the integral.

Using \refeq{eq:r_delta_epsilon}, we have
\begin{equation*}
\frac{\left(1-e^{-\epsilon}\right)^2}
  {1-2e^{-\epsilon}\cos\theta+e^{-2\epsilon}}
  = \frac{\left(1-e^{-\epsilon}\right)^2}{1-e^{-2\epsilon}}
    2\pi\delta_\epsilon(\theta)\text{.}
\end{equation*}

On substitution into \refeq{eq:r_lim_Q_exists}, we obtain

\begin{equation*}
\frac{\left(1-e^{-\epsilon}\right)^2}{1-e^{-2\epsilon}} 2\pi
  \int_{-\alpha}^\alpha \delta_\epsilon(\theta)
    \frac{m_y(d\theta)}{2(1-\cos\theta)}
  = \frac{\left(1-e^{-\epsilon}\right)^2}{1-e^{-2\epsilon}} 2\pi
       \int_{-\alpha}^\alpha \frac{d\Theta_\epsilon}{d\theta}
       \frac{g_y(\theta)}{2(1-\cos\theta)} d\theta\text{.}
\end{equation*}
The function
$\Theta_\epsilon(\theta) = \int \delta_\epsilon(\theta')d\theta'$ is the
step function in the
$\epsilon\rightarrow 0$ limit. For non--zero $\epsilon$ it is positive,
monotonic, increasing and bounded by unity. As $\theta\notin N$ we have
also written $m_y(d\theta) = g_y(\theta)d\theta$ for some well behaved
positive  function $g_y(\theta)$. By integration by parts (see
\cite{jeffreys}, p.~32 for existence conditions, which are satisfied) we
obtain
\begin{equation*}
\frac{\left(1-e^{-\epsilon}\right)^2}{1-e^{-2\epsilon}} 2\pi \left\{
  \left[ \Theta_\epsilon(\theta)
    \frac{g_y(\theta)}{2(1-\cos\theta)}\right]^\alpha_{-\alpha} -
  \int_{-\alpha}^\alpha \Theta_\epsilon(\theta)
    \frac{d}{d\theta}\frac{g_y(\theta)}{2(1-\cos\theta)} d\theta
    \right\}\text{.}
\end{equation*}
The first term within the curly braces is clearly some finite value. The
second term is less than
\begin{equation*}
\int_{-\alpha}^\alpha
  \frac{d}{d\theta} \frac{g_y(\theta)}{2(1-\cos\theta)} d\theta
 = \left[ \frac{g_y(\theta)}{2(1-\cos\theta)}\right]^\alpha_{-\alpha}
\end{equation*}
from the properties of the $\Theta_\epsilon$ function mentioned above. As
with the first term, it is clearly some finite value. Noting that
\begin{equation*}
\underset{\epsilon\downarrow 0}{\lim}
  \frac{\left(1-e^{-\epsilon}\right)^2}{1-e^{-2\epsilon}} = 0
\end{equation*}
we see that part (\ref{prop:r_3b}) follows.\proofend


\begin{thrm}
\label{thrm:r_4}
Let A be bounded and $U$-finite on $J$, with $R(A^*)$ cyclic for $U$. Let
$W$ be bounded and self-adjoint on $\mathcal{K}$, and define the Floquet
operator
\begin{equation*}
V = e^{iA^*WA/\hbar}U\text{.}
\end{equation*}
Assume that for $|z|  \neq 1$, $Q(z)$ is
compact, and that $Q\left(e^{\pm i(\theta \pm i\epsilon)}\right)$ converges
to $Q\left(e^{\pm i\theta}\right)$ in operator norm as $\epsilon\to 0$ for
a.e.\ $\theta$ in $J$. Define the set
\begin{equation*}
M(U,A,J) = \{\theta\in J : Q\left(e^{\pm i(\theta \pm i0)}\right)
 \text{ does not exist in norm} \}\text{.}
\end{equation*}

Then
\begin{enumerate}
\item $V$ has no absolutely continuous spectrum in $J$, and
  \label{thrm:r_4a}
\item the singular continuous part of the spectrum of $V$ in $J$ is
supported by the set $N(U,A,J) \cup M(U,A,J)$.
  \label{thrm:r_4b}
\end{enumerate}
\end{thrm}


\emph{Proof.} (\ref{thrm:r_4a}) 
For convenience, we write the Floquet operator as
\begin{equation*}
V = (1+A^*ZA)U
\end{equation*}
where $Z$ is defined appropriately by requiring\footnote{For the rank-N
perturbation case where
$W=\sum_{k=1}^N \lambda_k|\psi_k\rangle\langle\psi_k|$ and
$A = \sum_{k=1}^N |\psi_k\rangle\langle\psi_k|$, we have
$Z= \sum_{k=1}^N (\exp(i\lambda_k / \hbar)-1)|\psi_k\rangle\langle\psi_k|$.}
$\exp(iA^*WA / \hbar) = 1+A^*ZA$.
Noting \refeq{eq:r_Fdef} and \refeq{eq:r_QfromF} allows us to define
\begin{equation*}
\begin{split}
Q_1\left(e^{i\theta}\right) &= AF(\theta;V)A^* \\
  &= A\left(1-Ve^{i\theta}\right)^{-1}A^*\text{.}
\end{split}
\end{equation*}
Consider some vector $y' \in \mathcal{H}$. 
$Ay'= x \in \mathcal{K}$ is defined for such $y'$. $A^*x = y''$ is some
vector in $\mathcal{H}$. The cyclicity of $R(A^*)$ means that action with
linear combinations of powers of $U$ on $y''$ allows us to obtain any
$y \in \mathcal{H}$, our original $y'$ being one of them. Thus, we have a
construction of $A^{-1}$, namely, operation with $A^{*}$ followed by the
linear combination of powers of $U$. As $y'$ was arbitrary, $A^{-1}$ exists
for all $y \in \mathcal{H}$. This allows us to introduce $I=A^{-1}A$ in what
follows\footnote{The particular choice of $A$ as a projection in
\refeq{eq:r_rankNpert} does not have an inverse, but we will see in
\refsec{sec:r_finite} that we can define a subspace of $\mathcal{H}$ on which
$R(A^*)$ is cyclic, and apply this theorem.}.

We now proceed by use of the resolvent equation
\begin{equation}
\begin{split}
\label{eq:r_q1-q}
Q_1 &- Q \\
  &=
  A\left\{\frac{1}{1-Ve^{i\theta}} - \frac{1}{1-Ue^{i\theta}}\right\}A^* \\
  &= A\left\{\frac{1}{1-Ve^{i\theta}}\left(A^*ZAUe^{i\theta}\right)
    \frac{1}{1-Ue^{i\theta}}\right\}A^* \\
  &= Q_1\left(e^{i\theta}\right)ZAUA^{-1}
    e^{i\theta}Q\left(e^{i\theta}\right)\text{.}
\end{split}
\end{equation}
Thus, briefly using $L = ZAUA^{-1}e^{i\theta}$ for clarity, we have
\begin{alignat}{2}
\label{eq:r_q1_inverse_q}
&&LQ_1 - LQ &= LQ_1LQ \notag \\
&\Rightarrow& (1+LQ_1)(1-LQ) &= 1 \notag \\
&\Rightarrow& 1+e^{i\theta}ZAUA^{-1}Q_1\left(e^{i\theta}\right)
  &= \notag \\
&& \bigl[1-e^{i\theta}ZAU&A^{-1}Q\bigl(e^{i\theta}\bigr)\bigr]^{-1}\text{.}
\end{alignat}
Denote by $N$ and $M$ the sets $N(U,A,J)$ and $M(U,A,J)$. If
$\theta \in (J \sim N) \sim M$, i.e.,\ $\theta \in J \cap N^c \cap M^c$,
and $1-e^{i\theta}ZAUA^{-1}Q\left(e^{i\theta}\right)$ is not invertible,
then the compactness of $-LQ\left(e^{i\theta}\right)$ (which follows from the
compactness of $Q\left(e^{i(\theta+i\epsilon)}\right)$, the norm
convergence of $Q\left(e^{i(\theta+i\epsilon)}\right)$ and Theorem~VI.12 in
\cite{Reed1}) allows us to use the Fredholm Alternative
(\cite{Reed1}, p.~201, Theorem~VI.14) to assert that
\begin{equation*}
\exists x \in \mathcal{K},\text{ s.t. }
\left[1-e^{i\theta}ZAUA^{-1}Q\left(e^{i\theta}\right)\right]x = 0\text{.}
\end{equation*}
That is, there is some vector $x\in\mathcal{K}$ which satisfies the
equation
\begin{equation}
\label{eq:r_x_exists}
x - e^{i\theta}ZAUA^{-1}A\left(1-Ue^{i\theta}\right)^{-1}A^*x = 0\text{.}
\end{equation}
As $\theta \in J \sim N$, by Proposition~\ref{prop:r_3}
$y=A^*x \in D\left[\left(1-Ue^{i\theta}\right)^{-1}\right]$ so define $\phi$ as
\begin{equation}
\label{eq:r_phi}
\phi = \left(1-Ue^{i\theta}\right)^{-1}A^*x\text{.}
\end{equation}
$\phi$ is a well defined vector on $\mathcal{H}$ and we have
\begin{equation*}
x = e^{i\theta}ZAU\phi\text{.}
\end{equation*}
By \refeq{eq:r_phi}, $x \neq 0$ implies $\phi \neq 0$, so we have
\begin{alignat}{2}
&&\qquad\left(1-Ue^{i\theta}\right)\phi = A^*x &=
 e^{i\theta}A^*ZAU\phi \notag \\
&\text{or}&\qquad V\phi &=e^{-i\theta}\phi\text{.} \label{eq:r_f_e_value}
\end{alignat}
We conclude that $e^{-i\theta}\in\sigma_p(V)$.

The multiplicity of the eigenvalue is given by the dimension of the kernel
of $1-e^{i\theta}ZAUA^{-1}Q$, which is finite by the
compactness of $Q$ and Theorem~4.25 of \cite{RudinFA}.

Therefore, if $\theta \in J \sim (N \cup M \cup \sigma_p(V))$, which is a
set of full Lebesgue measure\footnote{That the set $M$ has
measure zero is a consequence of Lemma~\ref{lemma:r_5}.}, then the vector
\begin{equation}
\begin{split}
\label{eq:r_xeps}
x(\epsilon)
  &= \left[1+e^{i(\theta+i\epsilon)}ZAUA^{-1}Q_1\left(e^{i(\theta+i\epsilon)}
    \right)\right]x \\
  &\equiv \left[1+L_+Q_1\left(e^{i\theta_+}\right)\right]x
\end{split}
\end{equation}
must be bounded in norm as $\epsilon\to 0$ because we have just seen
that if it is unbounded, we have an eigenvalue of the operator $V$. For
$y=A^*x \in R(A^*)$, the absolutely continuous spectrum, $m_y^{ac}$ of $V$ is
the limit of
\begin{equation*}
\left\langle\delta_\epsilon\left(1-Ve^{i\theta}\right)y,y\right\rangle
  = \left\langle A \delta_\epsilon\left(1-Ve^{i\theta}\right)A^*x,x
  \right\rangle\text{.}
\end{equation*}
Our aim is to show that this is zero for all $y\in\mathcal{H}$. We define
\begin{align}
F_1(\theta) &= \left(1-Ve^{i\theta}\right)^{-1} \\
F(\theta) &= \left(1-Ue^{i\theta}\right)^{-1}
\end{align}
and in a similar fashion to \refeq{eq:r_q1-q} and \refeq{eq:r_q1_inverse_q},
obtain
\begin{equation}
\label{eq:r_F1resolvent}
F_1(\theta) = F(\theta)\left[1+ (V-U)e^{i\theta}F_1(\theta)\right]
\end{equation}
and
\begin{equation*}
\left(1+(V-U)e^{i\theta}F_1(\theta)\right)
  = \left(1-(V-U)e^{i\theta}F(\theta)\right)^{-1}\text{.}
\end{equation*}
Writing $X=V-U$, on substituting \refeq{eq:r_F1resolvent} into our expression
for the $\delta$-function \refeq{eq:r_delta_resolvent} we obtain
\begin{equation*}
\begin{split}
2\pi \delta_\epsilon\left(1-Ve^{i\theta}\right)
  &= \left(1-e^{-2\epsilon}\right)
    F_1^*(\theta_+)F_1(\theta_+)\\	
  &= \left[1+e^{i\theta_+}XF_1(\theta_+)\right]^*
    2\pi \delta_\epsilon\left(1-Ue^{i\theta}\right)
    \left[1+e^{i\theta_+}XF_1(\theta_+)\right]\text{.}
\end{split}
\end{equation*}
Substitution of \refeq{eq:r_QfromF} and noting that
\begin{equation*}
X = V - U = (1+A^*ZA)U - U = A^*ZAU
\end{equation*}
gives us
\begin{equation*}
\begin{split}
A&\delta_\epsilon\left(1-Ve^{i\theta}\right) A^* \\
  &= A\left[1+Xe^{i\theta_+}F_1(\theta_+)\right]^*
    \delta_\epsilon\left(1-Ue^{i\theta}\right)
    \left[1+Xe^{i\theta_+}F_1(\theta_+)\right]A^* \\
  &= \left[1 + L_+Q_1(\theta_+)\right]^*
    A\delta_\epsilon\left(1-Ue^{i\theta}\right)A^*
    \left[1 + L_+Q_1(\theta_+)\right]\text{.}
\end{split}
\end{equation*}
The absolutely continuous spectrum, $m_y^{ac}$ of $V$ is the limit of
\begin{equation}
\begin{split}
\label{eq:r_V_ac}
\bigl\langle &A\delta_\epsilon\left(1-Ve^{i\theta}\right)A^*x,x
  \bigr\rangle \\
  &= \left\langle\left[1+L_+Q_1(\theta_+)\right]^*
    A\delta_\epsilon\left(1-Ue^{i\theta}\right)A^*
    \left[1+L_+Q_1(\theta_+)\right]x,x\right\rangle \\
  &= \left\langle A\delta_\epsilon\left(1-Ue^{i\theta}\right)A^*
    x(\epsilon),x(\epsilon)\right\rangle \\
  &= \frac{\epsilon(1-\epsilon)}{\pi}
    \left\langle G_\epsilon(\theta;U,A)x(\epsilon),x(\epsilon)\right\rangle
\end{split}
\end{equation}
which tends to zero as $\epsilon\to 0$ if both $G_\epsilon(\theta;U,A)$
and $x(\epsilon)$ are bounded.
$G_\epsilon(\theta;U,A)$ is bounded as we have $\theta \in J \sim N$ and
$x(\epsilon)$ is bounded by \refeq{eq:r_xeps}.

Part (\ref{thrm:r_4a}) follows since $R(A^*)$ cyclic for $U$ implies that
$R(A^*)$ is cyclic for $V$.
\vspace{\baselineskip}  


(\ref{thrm:r_4b}) Let $N_1=N(V,A,J)$. We have just shown that
$\theta \in J \sim (N \cup M \cup \sigma_p(V))$ implies that
\begin{equation}
\label{eq:r_g_epsilon_v}
\frac{\epsilon}{\pi}\langle G_\epsilon(\theta;V,A)
  x(\epsilon),x(\epsilon)\rangle \to 0
\end{equation}
and therefore
\begin{equation}
\label{eq:r_delta_v}
\langle \delta_\epsilon\left(1-Ve^{i\theta}\right)y,y\rangle \to 0\text{.}
\end{equation}
If we can infer the strong limit from this weak limit then we
have established that $\theta \notin N_1$. We use the result that if
$x_n\overset{w}{\rightarrow} x$ and
$\lpar x_n \rpar \rightarrow \lpar x \rpar$, then
$x_n\overset{s}{\rightarrow} x$
(\cite{Bachman}, p 244). Writing $G_\epsilon$ and $G$ for
$G_\epsilon(\theta;V,A)$ and $G(\theta;V,A)$, and $F_\epsilon$ and $F$ for
$F(\theta_+;V)$ and $F(\theta;V)$, consider
\begin{equation*}
\begin{split}
\bigl| \lpar G_\epsilon &x \rpar^2 - \lpar G x \rpar^2 \bigr| \\
  &= \left| \langle (G_\epsilon^2 - G^2)x, x \rangle \right| \\
  &= \left| \langle A\left\{\left(F^*_\epsilon F_\epsilon - F^*F\right)
     A^*AF^*_\epsilon F_\epsilon + F^*FA^*A
     \left(F^*_\epsilon F_\epsilon - F^*F\right)
     \right\}A^* x, x \rangle \right|\text{.}
\end{split}
\end{equation*}
If $A$, $F_\epsilon$ and $F$ are bounded operators, then if
$F^*_\epsilon F_\epsilon - F^*F$ tends to zero as $\epsilon\rightarrow 0$
we can conclude that the strong limit exists. A short calculation
shows that
\begin{equation*}
F^*_\epsilon F_\epsilon - F^*F
  = \left[\left(1-e^{-2\epsilon}\right)
    - \left(1-e^{-\epsilon}\right)
    \left(Ue^{i\theta} + U^*e^{-i\theta}\right) \right]
    F^*_\epsilon F_\epsilon F^* F
\end{equation*}
which trivially tends to zero as $\epsilon\rightarrow 0$ given the
boundedness of $F_\epsilon$ and $F$. Finally, $A$ is bounded by assumption
and \refeq{eq:r_xeps} shows that $Q_1(\theta_+)$ is a bounded operator as
$\epsilon\rightarrow 0$ and thus both $F_\epsilon$ and $F$ are bounded.

Moving on from \refeq{eq:r_delta_v}, we have now established that
$N_1 \subset N \cup M \cup \sigma_p(V)$ so $N_1$ must have measure zero,
again remembering that we need Lemma~\ref{lemma:r_5} below to prove that $M$
has measure zero. By Theorem~\ref{thrm:r_2}, $N_1$ supports the singular
spectrum of $V$. That is,
\begin{equation*}
m^s\left(N_1^\text{c}\right) = 0
\end{equation*}
where the set $N_1^\text{c}$ is the complement of $N_1$.
As the measure is positive and $m^s = m^{sc} + m^p$, we know that
\begin{equation*}
m^{sc}\left(N_1^\text{c}\right) = 0\text{.}
\end{equation*}
Trivially, $(N \cup M)\sim \sigma_p(V)$ contains $N_1 \sim \sigma_p(V)$.
Thus
\begin{equation*}
\begin{split}
m^{sc}\left(\left[N_1 \cap \sigma_p(V)^\text{c}\right]^\text{c}\right)
  &= m^{sc}\left(N_1^\text{c} \cup \sigma_p(V)\right) \\
  &= m^{sc}\left(N_1^\text{c}\right) + m^{sc}\left(\sigma_p(V)\right) \\
  &= 0 + 0 \\
  &= 0
\end{split}
\end{equation*}
as the (continuous) measure of single points is zero.

The set $N \cup M \cap \sigma_p(V)^\text{c}$ must support
$m^{sc}$ as $N_1 \cap \sigma_p(V)^\text{c}$ is a subset. Therefore
\begin{equation*}
m^{sc}\left(\left[N \cup M \cap \sigma_p(V)^\text{c}\right]^\text{c}\right)
  = 0\text{.}
\end{equation*}
This equals
\begin{equation*}
\begin{split}
 m^{sc}\left(\left[N \cup M\right]^\text{c} \cup \sigma_p(V)\right)
  &= m^{sc}\left(\left[N \cup M\right]^\text{c}\right) +
      m^{sc}\left(\sigma_p(V)\right) \\
 &= m^{sc}\left(\left[N \cup M\right]^\text{c}\right)
\end{split}
\end{equation*}
so we conclude that the set $N \cup M$ supports the singular continuous
part of the spectrum.\proofend


Theorem~\ref{thrm:r_4} has shown us that $V$ has an empty absolutely
continuous component, and that the singular continuous component is
supported by the set $N \cup M$, which is independent of $\lambda$.
We know that $N$ has measure zero, and
Lemma~\ref{lemma:r_5} below shows us that $M$ also has measure zero. This
will allow us to apply Theorem~\ref{thrm:r_6} to show that the singular
continuous spectrum of $V$ is also empty. Thus, with both the a.c.\ and
s.c.\ spectra empty, we can conclude that $V$ must have pure point spectrum.


\begin{lemma}
\label{lemma:r_5}
Let $Q(z)$ be a trace class valued analytic function inside the complex
unit circle, with $|z|<1$. Then for a.e.\ $\theta$
\begin{equation*}
\underset{\epsilon\downarrow 0}{\lim} Q\left(e^{i(\theta +i\epsilon)}\right)
  \equiv
  Q\left(e^{i(\theta+i0)}\right)
\end{equation*}
exists in Hilbert Schmidt norm.
\end{lemma}


\emph{Proof.}
We parallel the proof of de Branges theorem (see \cite{deBranges} and pages
149-150 in \cite{Kato71}). Consider
\begin{equation*}
\begin{split}
Q\left(e^{i(\theta+i\epsilon)}\right)
  &+ Q^*\left(e^{i(\theta+i\epsilon)}\right) \\
 &= \int A^*\left\{\frac{1}{1-e^{-i(\theta'-\theta)}e^{-\epsilon}}
   + \frac{1}{1-e^{i(\theta'-\theta)}e^{-\epsilon}} \right\} AE(d\theta') \\
 &= \int A^*\left\{\frac{2\left(1-e^{-\epsilon}\cos(\theta'-\theta)\right)}
   {1+e^{-2\epsilon}-2e^{-\epsilon}\cos(\theta'-\theta)}\right\}
   AE(d\theta')\text{.}
\end{split}
\end{equation*}
The factor within the curly braces is greater than zero for all
$\theta',\theta$ and thus we have
\begin{equation*}
Q\left(e^{i(\theta+i\epsilon)}\right) + Q^*\left(e^{i(\theta+i\epsilon)}\right)
 \geq 0 \phantom{a}\forall \epsilon \geq 0\text{.}
\end{equation*}
Therefore, following de Branges,
\begin{equation*}
\begin{split}
\left| \det \left(1+Q\left(e^{i(\theta+i\epsilon)}\right)\right) \right|^2
 &\geq \det \left( 1+Q^*\left(e^{i(\theta+i\epsilon)}\right)
   Q\left(e^{i(\theta+i\epsilon)}\right) \right) \\
 &= \prod\left(1 + |\alpha_n|^2\right) \\
 &\geq
    \begin{cases}
      \sum|\alpha_n|^2 =
        \lpar Q\left(e^{i(\theta+i\epsilon)}\right) \rpar^2_{H.S.} \\
      1\text{.}
    \end{cases}
\end{split}
\end{equation*}
$\{\alpha_n\}$ are the eigenvalues of $Q\left(e^{i(\theta+i\epsilon)}\right)$.
From the two bounds on
$\left|\det\left(1+Q\left(e^{i(\theta+i\epsilon)}\right)\right)\right|$
above, we obtain
\begin{equation*}
\left|\left| \frac{Q\left(e^{i(\theta+i\epsilon)}\right)}
  {\det \left(1+Q\left(e^{i(\theta+i\epsilon)}\right)\right)}
  \right|\right|_{H.S.} \leq 1 \text{ and }
\left| \frac{1}{\det \left(1+Q\left(e^{i(\theta+i\epsilon)}\right)\right)}
  \right| \leq 1\text{.}
\end{equation*}
The definition of an analytic operator (\cite{Reed1}, p.~189) implies the
analyticity of the eigenvalues, and thus the operations of taking the
determinant and the Hilbert Schmidt norm are analytic. Hence, both functions
above are analytic and bounded within the complex unit circle
($\epsilon > 0$). Application of Fatou's theorem (\cite{Dienes}, p.~454)
establishes the existence in the limit as $\epsilon\rightarrow 0$ and hence
both functions exist on the boundary almost everywhere. Taking the quotient
we establish the existence of $Q\left(e^{i(\theta+i0)}\right)$ in the
Hilbert Schmidt norm.\proofend


Let $(\Omega,\mu)$ be a separable measure space, and
\begin{equation*}
V(\lambda) = \int e^{-i\theta} E_\lambda(d\theta)
\end{equation*}
a measurable family of unitary operators on $\mathcal{H}$. We denote by
\begin{equation*}
\mathbb{V} = \int e^{-i\theta} \mathbb{E}(d\theta)
\end{equation*}
the multiplication operator
\begin{equation*}
(\mathbb{V}u)(\lambda) = V(\lambda)u(\lambda)
\end{equation*}
on $L^2(\Omega,\mu;\mathcal{H})$, where
$u(\lambda) \in L^2(\Omega,\mu;\mathcal{H})$.

A vector $u(\lambda)$ is an element of $L^2(\Omega,\mu;\mathcal{H})$ if,
for $u(\lambda)\in\mathcal{H}$,
\begin{equation*}
\int_{-\infty}^\infty \lpar u(\lambda) \rpar ^2 d\mu < \infty\text{.}
\end{equation*}

It is important to note the difference between $V(\lambda)$ acting on
$\mathcal{H}$ and $\mathbb{V}$ acting on $L^2(\Omega,\mu;\mathcal{H})$. To
obtain our goal of showing that for a.e.\ $\lambda$, $V(\lambda)$ has a pure
point spectrum, we must show that $\mathbb{V}$ is absolutely continuous as
a function of $\lambda$ on the space $L^2(\Omega,\mu;\mathcal{H})$.

Theorem~\ref{thrm:r_6} is taken directly from \cite{Howland87}. The proof given
is, apart from some small notational changes,
identical to that in \cite{Howland87}. Due to a number of typographical
errors however, we have reproduced the proof here for reference and clarity.


\begin{thrm}
\label{thrm:r_6}
Let $\mathbb{V}$ be absolutely continuous on $L^2(\Omega,\mu;\mathcal{H})$,
and assume that there is a fixed set $S$ of Lebesgue measure zero which
supports the singular continuous spectrum of $V(\lambda)$ in the interval $J$
for $\mu$-a.e.\ $\lambda$. Then $V(\lambda)$ has no singular continuous
spectrum in $J$ for $\mu$-a.e.\ $\lambda$.
\end{thrm}
 

\emph{Proof.} For fixed $x\in\mathcal{H}$, and any measurable subset
$\Gamma$ of $\Omega$, let
$u(\lambda) = \chi_{\text{\tiny\itshape $\Gamma$}}(\lambda)x$ be a vector
in $L^2(\Omega,\mu;\mathcal{H})$. Then
\begin{equation*}
\begin{split}
\int_\Gamma \left| E_\lambda^{sc}[J]x\right|^2\mu(d\lambda)
  &\leq \int_\Gamma \left| E_\lambda[S]x\right|^2\mu(d\lambda) \\
  &= \int \left| E_\lambda[S]u(\lambda)\right|^2\mu(d\lambda) \\
  &= \int \left| \mathbb{E}[S]u(\lambda) \right|^2\mu(d\lambda) \\
  &= \;\lpar \mathbb{E}[S]u(\lambda)\rpar^2 \;= 0\text{.}
\end{split}
\end{equation*}

$\int_\Gamma \left| E_\lambda^{sc}[J]x\right|^2\mu(d\lambda) = 0$ implies
that $\left|E_\lambda^{sc}[J]x\right|^2 = 0$ for $\mu$-a.e.\ $\lambda$. Thus
\begin{equation*}
E_\lambda^{sc}[J]x = 0
\end{equation*}
for every $x\in\mathcal{H}$.\proofend


The application of Theorem~\ref{thrm:r_6} relies on finding a fixed set $S$
of measure zero which supports the singularly continuous spectrum.
$S = N \cup M$ is sufficient.

We have now established all the basic requirements for $V$ to be pure
point, given $U$ pure point. They are now combined to produce the main
theorem of the paper. There is still quite a lot of manipulation to satisfy
the condition $\mathbb{V}$ a.c.\ on $L^2(\mathbb{R};\mathcal{H})$ of
Theorem~\ref{thrm:r_6}, and this will be the focus for the remainder of
\refsec{sec:r_spectra} and \refsec{sec:r_pk}.


\begin{thrm}
\label{thrm:r_7}
Let $U$ and $A$ satisfy the hypotheses of Theorem~\ref{thrm:r_4} and define
for $\lambda \in \mathbb{R}$
\begin{equation*}
V(\lambda) = e^{i\lambda A^*A / \hbar}U\text{.}
\end{equation*}
Then $V(\lambda)$ is pure point in $J$ for a.e.\ $\lambda$.
\end{thrm}


\emph{Proof.} By Theorem~\ref{thrm:r_4}, with $W=\lambda I$,
$V(\lambda)$ has no absolutely continuous spectrum in $J$, and its singularly
continuous spectrum is supported on the fixed set $S = N \cup M$. Application
of Lemma~\ref{lemma:r_5} shows that $S$ is of measure zero.
Theorem~\ref{thrm:r_6} applies and shows that the singular continuous
spectrum is empty,
if we can show that $\mathbb{V}$ is absolutely continuous on
$L^2(\mathbb{R};\mathcal{H})$. We show this in the following sections.

As we have shown that both the absolutely continuous and singular continuous
parts of the spectrum are empty, we conclude that $V(\lambda)$ is pure point
for a.e.\ $\lambda \in \mathbb{R}$.\proofend


To show that $\mathbb{V}$ is a.c., we apply a modified version of the
Putnam--Kato theorem which is proved in \refsec{sec:r_pk}. The unitary
Putnam--Kato theorem is

\vspace{\baselineskip}
\noindent{\textbf\itshape Theorem~\ref{thrm:r_unitpk}}
{\itshape Let $V$ be unitary, and $D$ a self-adjoint bounded operator. If
$C=V[V^*,D] \geq 0$, then $V$ is absolutely continuous on $R(C^{1/2})$.
Hence, if $R(C^{1/2})$ is cyclic for $V$, then $V$ is absolutely
continuous on $\mathcal{H}$.}
\vspace{\baselineskip}  

We apply this theorem on the space $L^2(\mathbb{R};\mathcal{H})$. A naive
application to obtain the desired result is as follows. We slightly change
notation and explicitly include the $\lambda$ dependence of $W$ in our
definition of $V$. If we choose $\mathbb{D} = -i(d/d\lambda)$, with
$V = e^{i\lambda A^*WA}U$, we see that
\begin{equation*}
-i\frac{dV^*}{d\lambda} = -U^*A^*WA e^{-i\lambda A^*WA} = -V^*A^*WA
\end{equation*}
so that for some $u \in L^2(\mathbb{R};\mathcal{H})$
\begin{equation*}
\begin{split}
[\mathbb{V}^*,\mathbb{D}]u &= (\mathbb{V}^*\mathbb{D} -
  \mathbb{D}\mathbb{V}^*)u = -\mathbb{D}\mathbb{V}^*u \\
  &= i\frac{d}{d\lambda}(\mathbb{V}^*u) = \mathbb{V}^*A^*WAu\text{.}
\end{split}
\end{equation*}
Therefore,
\begin{equation*}
\mathbb{C} = \mathbb{V}[\mathbb{V}^*,\mathbb{D}] = A^*WA\text{.}
\end{equation*}
With $W=I$, we obtain $\mathbb{C} = A^*A \geq 0$ and thus
$R(\mathbb{C}^{1/2})=R(A^*)$ (see
the proof to Theorem~VI.9 in \cite{Reed1}) is cyclic for $V$. Hence,
$\mathbb{V}$ is a.c.\ and we satisfy all the requirements of
Theorem~\ref{thrm:r_7}.

The problem here is that $\mathbb{D}$ is not bounded, and boundedness of
$\mathbb{D}$ is essential in the proof of the Putnam--Kato theorem. We use
a similar technique as Howland \cite{Howland87} to overcome this issue. 

As the norm of $A^*A$ may be scaled however we like, we can rewrite $V$, for
real $t$ as
\begin{equation}
\label{eq:r_scaled_v}
V(t) = e^{ictA^*A}U
\end{equation}
for some real $c>0$.


\begin{prop}
\label{prop:r_8}
On $L^2(\mathbb{R};\mathcal{H})$, consider the unitary multiplication operator
$\mathbb{V}$, defined by
\begin{equation*}
\mathbb{V}u(t) = V(t)u(t) = e^{ictA^*A}Uu(t)
\end{equation*}
and the bounded self-adjoint operator $\mathbb{D}=-\arctan(p/2)$,
where $p=-id/dt$.
Then $C=\mathbb{V}[\mathbb{V}^*,D]$ is positive definite, and
$R\left(C^{1/2}\right)$ is cyclic for $\mathbb{V}$. Hence, the requirements
of Theorem~\ref{thrm:r_7} are fully satisfied.
\end{prop}


\emph{Proof.} The operator $\mathbb{D}$ on $L^2(\mathbb{R};\mathcal{H})$
is convolution by the Fourier transform of $-\arctan(x/2)$ \cite{Howland87},
which is $i\pi t^{-1} e^{-2|t|}$ (\cite{Erdelyi}, p.~87, (3)). This is a
singular (principal value) integral operator, because $\arctan(p/2)$ does
not vanish at infinity. Thus, for $u(t) \in L^2(\mathbb{R};\mathcal{H})$,
\begin{equation*}
\mathbb{D}u(t) 
  = i\pi P \int_{-\infty}^{\infty} \frac{e^{-2|t-y|}}{t-y}u(y)dy
\end{equation*}
and
\begin{equation*}
[\mathbb{V}^*,\mathbb{D}]u(t) = i\pi P \int_{-\infty}^{\infty}
  e^{-2|t-y|}\frac{V^*(t)-V^*(y)}{t-y}u(y)dy
\end{equation*}
so
\begin{equation}
\begin{split}
\label{eq:r_C}
\mathbb{C}u(t) &= \mathbb{V}[\mathbb{V}^*,\mathbb{D}]u(t) \\
  &= i\pi P \int_{-\infty}^{\infty}
    e^{-2|t-y|}\frac{1-V(t)V^*(y)}{t-y}u(y)dy\text{.}
\end{split}
\end{equation}
Inserting expression \refeq{eq:r_scaled_v} for $V(t)$, we obtain
\begin{equation}
\begin{split}
\label{eq:r_pk_int}
\mathbb{C}u(t) &=i\pi \int_{-\infty}^{\infty}
    e^{-2|t-y|}\frac{1-e^{ic(t-y)A^*A}}{t-y}u(y)dy \\
  &=i\pi \int_{-\infty}^{\infty}
    e^{-2|t-y|}\frac{1-\cos\left(A^*Ac(t-y)\right)
    -i\sin\left(A^*Ac(t-y)\right)}{t-y}u(y)dy\text{.}
\end{split}
\end{equation}
Note that this is no longer a singular integral. To show that $\mathbb{C}$
is positive, we must show that
\begin{equation*}
(u(t),\mathbb{C}u(t)) > 0
  \phantom{aa}\forall\phantom{a} u(t)\in L^2(\mathbb{R};\mathcal{H}).
\end{equation*}
Note that the inner product on $L^2(\mathbb{R};\mathcal{H})$ is given by
\begin{equation}
\label{eq:r_L_inner_prod}
(u(t),u'(t)) = \int_{-\infty}^\infty u^*(t)u'(t)dt\text{.}
\end{equation}
We now write our operator $A$ in terms of its spectral
components. Note that here $\lambda$ decomposes $A$ and bears no relation
to the strength parameter used at other stages in this paper. When required
for clarity, we write $\int_\lambda$ to identify the integral over the
variable $\lambda$.
\begin{equation*}
A = \int \lambda E(d\lambda)\text{.}
\end{equation*}
A general vector $u(t)$ may be written
\begin{equation*}
u(t) = \int E(d\lambda)u(t)\text{.}
\end{equation*}
Then
\begin{equation*}
f(A)u(t) = \int f(\lambda) E(d\lambda)u(t)
\end{equation*}
which implies that we may rewrite \refeq{eq:r_pk_int} as
\begin{equation*}
\begin{split}
\mathbb{C}u(t)
  &= i\pi \int_{-\infty}^\infty dy \int_\lambda
    e^{-2|t-y|}\frac{1-e^{ic(t-y)|\lambda|^2}}{t-y}
    E(d\lambda)u(y) \\
  &= \int_{-\infty}^\infty dy \int_\lambda \phi_\lambda(t-y)
  E(d\lambda)u(y) \\
  &= \int_\lambda E(d\lambda)\mathcal{C}_\lambda(t)
\end{split}
\end{equation*}
where
\begin{equation*}
\mathcal{C}_\lambda(t)
  = \int_{-\infty}^\infty dy \phi_\lambda(t-y)u(y)
\end{equation*}
and we have defined the new function
\begin{equation*}
\phi_\lambda(t) = i\pi e^{-2|t|}t^{-1}\left(1-e^{ict|\lambda|^2}
  \right)\text{.}
\end{equation*}
By the convolution theorem, note that
\begin{equation*}
\tilde{\mathcal{C}}_\lambda(\omega)
  = \tilde{\phi}_\lambda(\omega)\tilde{u}(\omega)
\end{equation*}
where the ``$\tilde{\phantom{a}}$'' indicates Fourier transform.

Using this decomposition of $u(t)$ and Parseval's theorem, we can now easily
write down $(u(t),\mathbb{C}u(t))$. We use $(x,y)_\mathcal{H}$ to indicate
the inner product on the Hilbert Space $\mathcal{H}$, reserving $(x,y)$ for
the inner product on $L^2(\mathbb{R};\mathcal{H})$ as in
\refeq{eq:r_L_inner_prod}.
\begin{equation*}
\begin{split}
\left(u(t),\mathbb{C}u(t)\right)
  &= \int_{-\infty}^\infty dt
    \left(u(t),\mathbb{C}u(t)\right)_{\mathcal{H}} \\
  &= \int_{-\infty}^\infty dt \left(u(t),
    \int_\lambda E(d\lambda)\mathcal{C}_\lambda(t)\right)_{\mathcal{H}} \\
  &= \int_{-\infty}^\infty dt \left(u(t), \int_\lambda E(d\lambda)
    \int \frac{d\omega}{2\pi}e^{i\omega t}\tilde{\mathcal{C}}_\lambda(\omega)
    \right)_{\mathcal{H}} \\
  &= \int_\lambda E(d\lambda) \int \frac{d\omega}{2\pi}
    \left|\tilde{u}_\lambda(\omega)\right|^2
    \tilde{\phi}_\lambda(\omega)\text{.}
\end{split}
\end{equation*}
We clearly see that if $\tilde{\phi}_\lambda(\omega)$ is positive for all
$\lambda$ then $\mathbb{C}$ will be positive.

In the following calculation we will find the need to bound $c|\lambda|^2$.
The restriction $0 \leq c|\lambda|^2\leq 1$ will be employed. We argue that
as $A^*A$ is a positive self-adjoint bounded operator we can restrict the
integral over $\lambda$ to (\cite{Riesz}, p.~262, 273)
\begin{equation}
A^*A = \int_{-\infty}^\infty |\lambda|^2 E(d\lambda)
  = \int_{m-0}^M |\lambda|^2 E(d\lambda)
\end{equation}
where $M$ is the least upper bound and $m$ the greatest lower bound of
$A^*A$. The norm of $A^*A$ is given by $\max(|m|,|M|)$. Thus, if we set
\begin{equation*}
c = \frac{1}{\lpar A^*A \rpar}
\end{equation*}
then we guarantee each $c|\lambda|^2$ to be less than unity.

Proceeding, the Fourier transform, $\tilde{\phi}_\lambda(\omega)$ of
\begin{equation}
\label{eq:r_phi_convolution}
\phi_\lambda(t) 
  = i\pi e^{-2|t|}t^{-1}\left[ 1-\cos ct|\lambda|^2 - i\sin ct|\lambda|^2
    \right]
\end{equation}
is now calculated. We split \refeq{eq:r_phi_convolution} into two parts.
\begin{align}
\label{eq:r_phi_1}
\phi_{\lambda1}(t) &= i\pi e^{-2|t|}t^{-1}\left[1-\cos ct|\lambda|^2\right] \\
\label{eq:r_phi_2}
\phi_{\lambda2}(t) &= \pi e^{-2|t|}t^{-1}\sin ct|\lambda|^2\text{.}
\end{align}
The Fourier transform of \refeq{eq:r_phi_1} is
\begin{equation*}
\begin{split}
\tilde{\phi}_{\lambda1}(\omega) &= i\pi\int_{-\infty}^{\infty}
  e^{-2|t|}t^{-1}(1-\cos ct|\lambda|^2)e^{-i\omega t} dt \\
  &= i\pi\Biggl[
    \int_0^{\infty} e^{-2t}t^{-1}(1-\cos ct|\lambda|^2)e^{-i\omega t} dt \\
  &\qquad  
    + \int_0^{\infty} e^{-2t}(-t^{-1})(1-\cos ct|\lambda|^2)e^{i\omega t} dt
    \Biggr]\text{.}
\end{split}
\end{equation*}
Using (\cite{Erdelyi}, p.~157, (59)), and setting
$S=c|\lambda|^2/(2+i\omega)$, we obtain
\begin{equation*}
\tilde{\phi}_{\lambda1}(\omega)
  = \frac{i\pi}{2}\log\left(\frac{1+S^2}{1+S^{*2}}\right)\text{.}
\end{equation*}
The logarithm of a complex number can in general be written as
\begin{equation*}
\log(z) = \log(|z|) + i\Arg z
\end{equation*}
so noting that $\left|(1+S^2)/(1+S^{*2})\right|=1$, we see that
\begin{equation*}
\begin{split}
\tilde{\phi}_{\lambda1}(\omega)
  &= -\frac{\pi}{2}\Arg\left(\frac{1+S^2}{1+S^{*2}}\right) \\
  &= -\pi\Arg\left(1+S^2\right)\text{.}
\end{split}
\end{equation*}
With $\kappa = c|\lambda|^2$, the real and imaginary parts of $1+S^2$ are
\begin{align*}
\Re\left(1+S^2\right)
  &= \frac{\left(4+\omega^2\right)^2 + \kappa^2\left(4-\omega^2\right)}
          {\left(4+\omega^2\right)^2} \\
\Im\left(1+S^2\right)
  &= \frac{-4\kappa^2\omega}{\left(4+\omega^2\right)^2}\text{.}
\end{align*}
With the restriction that $0 \leq \kappa \leq 1$, the real part is positive
for all $\omega$ and thus $\Arg(z) = \arctan(\Im z / \Re z)$. Thus,
\begin{equation*}
\tilde{\phi}_{\lambda1}(\omega)
  = -\pi\arctan\left(\frac{\Im\left(1+S^2\right)}{\Re\left(1+S^2\right)}
    \right)\text{.}
\end{equation*}
$\arctan(z)$ is the principal part of $\Arctan(z)$, with range
$-\pi/2 < \arctan(z) < \pi/2$. The Fourier transform of \refeq{eq:r_phi_2} is
similarly calculated using (\cite{Erdelyi}, p.~152, (16)), to be
\begin{equation*}
\begin{split}
\tilde{\phi}_{\lambda2}(\omega)
  &= \pi\left[\arctan S + \arctan S^* \right] \\
  &= \pi\left[\arctan\left(\frac{c|\lambda|^2}{2+i\omega}\right)
    + \arctan\left(\frac{c|\lambda|^2}{2-i\omega}\right)\right]\text{.}
\end{split}
\end{equation*}
Repeated application of the formula
$\arctan(z_1)+\arctan(z_2)=\arctan(z_1+z_2/1-z_1z_2)$, valid when
$z_1z_2 < 1$ (true for $0 \leq\kappa\leq 1$), yields\footnote{
This result is not valid for values of $\kappa$ larger than
around $2$, at which point the $\arctan$ addition formulas fail---this is
a moot point however, as we may trivially restrict $\kappa$ as already
explained.}
\begin{equation}
\begin{split}
\label{eq:r_Phi}
\tilde{\phi}_\lambda(\omega)
  &= \tilde{\phi}_{\lambda1}(\omega)+\tilde{\phi}_{\lambda2}(\omega) \\
  &= \pi\arctan\left(\frac{n(\omega,c|\lambda|^2)}
    {d(\omega,c|\lambda|^2)}\right)
\end{split}
\end{equation}
where
\begin{align}
\label{eq:r_numerator}
n(\omega,\kappa) &= 4\kappa\bigl[
  \left(4+\omega^2\right)^2 + \kappa\omega\left(4+\omega^2\right)
  + \kappa^2\left(4-\omega^2\right) - \kappa^3\omega \bigr] \\
\intertext{and}
d(\omega,\kappa) &= \left(4+\omega^2\right)^3
   -2\kappa^2\omega^2\left(4+\omega^2\right)
   - 16\kappa^3\omega - \kappa^4\left(4-\omega^2\right)\text{.}
\end{align}
One may easily confirm that for $0 \leq \kappa \leq 1$,
$n(\omega,\kappa)/d(\omega,\kappa)$ and hence $\tilde{\phi}_\lambda(\omega)$
is strictly positive by noting that there are four distinct regions of
interest for $\omega$, in which terms in $n$ and $d$ do not change sign.
\reftab{table:fourier} shows these regions and the sign of each term in
the region. Note that the global (positive and hence irrelevant) $\kappa$
factor from \refeq{eq:r_numerator} is dropped from the numerator for the
following discussion.
\begin{table}[H]
\begin{tabular}{|c|c|c|c|c|}\hline
$n(\omega,\kappa)=$ & $\left(4+\omega^2\right)^2$ &
  $+\kappa\omega\left(4+\omega^2\right)^2$ &
  $+\kappa^2\left(4-\omega^2\right)$ & $-\kappa^3\omega$ \\ \hline

$\omega < -2$ & +ve & -ve & -ve & +ve \\
$-2 < \omega < 0$ & +ve & -ve & +ve & +ve \\
$0 < \omega < 2$ & +ve & +ve & +ve & -ve \\
$\omega > 2$ & +ve & +ve & -ve & -ve \\ \hline \hline

$d(\omega,\kappa)=$ & $\left(4+\omega^2\right)^3$ &
  $-2\kappa^2\omega^2\left(4+\omega^2\right)$ &
  $-16\kappa^3\omega$ & $-\kappa^4\left(4-\omega^2\right)$ \\ \hline

$\omega < -2$ & +ve & -ve & +ve & +ve \\
$-2 < \omega < 0$ & +ve & -ve & +ve & -ve \\
$0 < \omega < 2$ & +ve & -ve & -ve & -ve \\
$\omega > 2$ & +ve & -ve & -ve & +ve \\ \hline
\end{tabular}
\caption{Sign of each term in the numerator $n(\omega,\kappa)$ and the
denominator $d(\omega,\kappa)$ of \refeq{eq:r_Phi}.}
\label{table:fourier}
\end{table}
For each row in the table, we simply need to show that the terms add to
produce a strictly positive number. First note that the first column for
both the numerator and denominator is independent of $\kappa$. To show the
positivity of each row, we set all positive $\kappa$-dependent terms to
zero and then take $\kappa=1$ for the negative terms to maximise their
contribution. Expanding out terms, it is then trivially seen in all cases
that the first column ($\left(4+\omega^2\right)^2$ for the numerator and
$\left(4+\omega^2\right)^3$ for the denominator) dominates. Thus, no row is
negative and we conclude that $\tilde{\phi}_\lambda$ is positive definite.

\begin{figure}[H]
\begin{center}
  \input{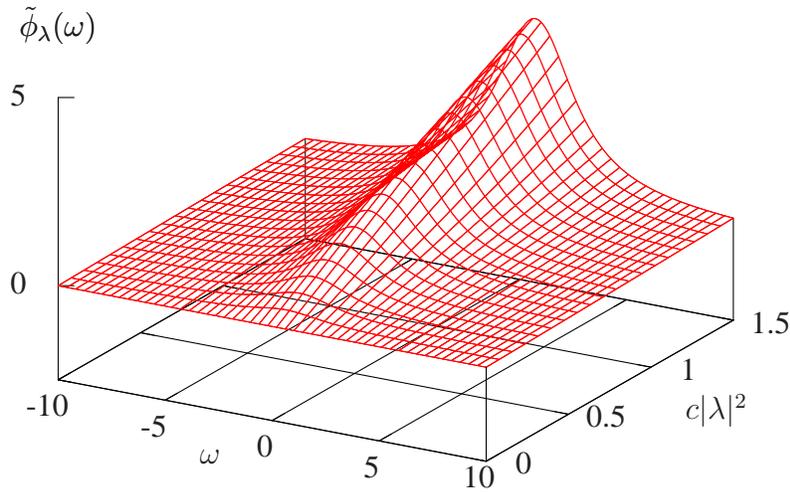}
  \caption{Plot of $\tilde{\phi}_\lambda(\omega)
  = \tilde{\phi}_{\lambda1}(\omega)
  + \tilde{\phi}_{\lambda2}(\omega)$, the Fourier transform of
  $\phi_\lambda(t) = i\pi e^{-2|t|}t^{-1}\left(1-\cos(ct|\lambda|^2)
  - i\sin(ct|\lambda|^2)\right)$. $\tilde{\phi}_\lambda(\omega)$ is strictly
  positive for all $\omega$ when $c|\lambda|^2 \leq 1$.}
  \label{fig:r_fourier}
\end{center}
\end{figure}

We have established that the Fourier transform of $\phi_\lambda$ is positive
definite for $c|\lambda|^2 \leq 1$. As a visual aid, \reffig{fig:r_fourier}
shows $\tilde{\phi}_\lambda(\omega)$. The positivity for
$c|\lambda|^2 \leq 1$ is clear.

Thus,
$\mathbb{C}$ is strictly positive and $\mathbb{V}$ is absolutely continuous
on $R(\mathbb{C}^{1/2})$. As $A^*A$ is a factor of $1-e^{ictA^*A}$
(i.e.,\ $A^*A$ is a factor of $\mathbb{C}$), $R(\mathbb{C}^{1/2}) = R(A^*)$.
Noting that $R(A^*)$ is cyclic for $U$ and hence cyclic for $V$, we conclude
that $R(\mathbb{C}^{1/2})$ is cyclic for $\mathbb{V}$. Thus, $\mathbb{V}$ is
absolutely continuous on $L^2(\mathbb{R};\mathcal{H})$.\proofend

We have now satisfied all the requirements of Theorem~\ref{thrm:r_7}.


\section{\label{sec:r_pk}The unitary Putnam--Kato theorem}

In this section, we will prove a modified version of the Putnam--Kato
theorem, as used in the preceding pages. The theorems and proofs follow a
similar argument to that of Reed and Simon (\cite{Reed4}, p.~157,
Theorem~XIII.28) and are motivated by the stroboscopic nature of the kicked
Hamiltonian.


\begin{defn}[V-Smooth]
\label{defn:vsmooth}
Let $V$ be a unitary operator. $A$ is $V$-smooth if and only if for all
$\phi \in \mathcal{H}$, $V(t)\phi\in D(A)$ for almost every
$t\in\mathbb{R}$ and for some constant $C$,
\begin{equation*}
\sum_n \lpar AV^n\phi \rpar^2 \;\leq C\lpar\phi\rpar^2\text{.}
\end{equation*}
\end{defn}


\begin{thrm}
\label{thrm:r_ranA_ac}
If $A$ is $V$-smooth, then
$\overline{R(A^*)} \subset \mathcal{H}_{ac}(V)$.
\end{thrm}


\emph{Proof.} Since $\mathcal{H}_{ac}(V)$ is closed, we need only show 
$R(A^*) \subset \mathcal{H}_{ac}(V)$. Let $\phi \in D(A^*)$,
$\psi = A^*\phi$, and let $d\mu_\psi$ be the spectral measure for $V$
associated with $\psi$. Define, for the period $T$ in \refeq{eq:r_hamiltonian}
\begin{equation}
\label{eq:r_f_n}
\mathcal{F}_n(T) = \frac{1}{\sqrt{2\pi}}\left(A^*\phi,[V(T)]^n\psi
  \right)\text{.}
\end{equation}
We calculate, droppoing the $T$ for clarity,
\begin{equation*}
\begin{split}
|\mathcal{F}_n|
  &= \frac{1}{\sqrt{2\pi}}\left| \left(\phi, AV^n\psi\right) \right| \\
  &\leq \frac{1}{\sqrt{2\pi}}\lpar\phi\rpar\lpar AV^n\psi \rpar\text{.}
\end{split}
\end{equation*}
Because $A$ is $V$-smooth, we see that
\begin{equation*}
\begin{split}
\sum_n |\mathcal{F}_n|^2
  &\leq \frac{1}{2\pi} \lpar\phi\rpar^2
    \sum_n \lpar AV^n\psi \rpar^2 \\
  &\leq \frac{C}{2\pi} \lpar\phi\rpar^2 \lpar\psi\rpar^2 \\
  &< \infty\text{.}
\end{split}
\end{equation*}
So $\mathcal{F}_n \in L^2(\mathbb{R})$. By the Riesz--Fischer theorem
(\cite{RudinRC}, p.~96--7, 4.26 Fourier Series),
$\mathcal{F}(\theta) = \frac{1}{\sqrt{2\pi}}
  \sum_n \mathcal{F}_n e^{-in\theta} \in L^2$. 

The spectral resolution of $V[T]$ is
\begin{equation*}
V[T] = \int_0^{2\pi} e^{i\theta} dE_T(\theta)
\end{equation*}
so we have
\begin{equation*}
(V[T])^n = \int_0^{2\pi} e^{in\theta} dE_T(\theta)\text{.}
\end{equation*}
Therefore, from \refeq{eq:r_f_n} we obtain
\begin{equation*}
\begin{split}
\mathcal{F}_n &= \frac{1}{\sqrt{2\pi}}\int_0^{2\pi}
  \left(A^*\phi, e^{in\theta} dE_T(\theta)\psi\right) \\
  &= \frac{1}{\sqrt{2\pi}}\int_0^{2\pi}
    e^{in\theta}\left(\psi, dE_T(\theta)\psi\right) \\
  &= \frac{1}{\sqrt{2\pi}}\int_0^{2\pi}
    e^{in\theta} d\mu_\psi(\theta)\text{.}
\end{split}
\end{equation*}
Using the inverse of the expression above for $\mathcal{F}(\theta)$ gives
\begin{equation*}
  \mathcal{F}_n = \frac{1}{\sqrt{2\pi}}\int_0^{2\pi}
    e^{in\theta} \mathcal{F}(\theta)d\theta\text{.}
\end{equation*}
As we have just shown that $\mathcal{F}(\theta) \in L^2$,
$d\mu_\psi(\theta) = \mathcal{F}(\theta)d\theta$ is absolutely continuous,
which implies that $\psi \in R(A^*)$ is in $\mathcal{H}_{ac}(V)$ and so
$\overline{R(A^*)} \subset \mathcal{H}_{ac}(V)$.\proofend


\begin{thrm}[Unitary Putnam--Kato theorem]
\label{thrm:r_unitpk}
Let $V$ be a unitary operator, and $A$ a self-adjoint bounded operator. If
$C=V[V^*,A] \geq 0$, then $V$ is absolutely continuous on $R(C^{1/2})$.
Hence, if $R(C^{1/2})$ is cyclic for $V$, then $V$ is absolutely
continuous.
\end{thrm}


\emph{Proof.} The discrete time-evolution of an operator $A$ is given by
\begin{equation*}
\mathcal{F}_n = V^{-n}AV^n\text{.}
\end{equation*}
We calculate
\begin{equation*}
\begin{split}
\mathcal{F}_n - \mathcal{F}_{n-1}
  &= V^{-n}V[V^*,A]V^n \\
  &\equiv G_n
\end{split}
\end{equation*}
so
\begin{equation*}
\begin{split}
\sum_{n=a}^{b}\left(\phi,G_n\phi\right)
  &= \sum_{n=a}^{b}\left(\phi,V^{-n}V[V^*,A]V^n\phi\right) \\
  &= \sum_{n=a}^{b}\left(V^n\phi,V[V^*,A]V^n\phi\right) \\
  &= \sum_{n=a}^{b}
    \left(C^{\frac{1}{2}}V^n\phi, C^{\frac{1}{2}}V^n\phi\right) \\
  &= \sum_{n=a}^{b}\lpar C^{\frac{1}{2}}V^n\phi\rpar^2
\end{split}
\end{equation*}
where $C = V[V^*,A]$. We also have
\begin{equation*}
\sum_{n=a}^{b}\left(\phi,G_n\phi\right) = 
  \left(\phi,V^{-b}AV^b\phi \right) -
  \left(\phi,V^{-(a-1)}AV^{(a-1)}\phi \right)\text{.}
\end{equation*}
Taking the modulus and using the Schwartz inequality, we
obtain
\begin{equation*}
\begin{split}
\sum_{n=a}^{b}\lpar C^{\frac{1}{2}}V^n\phi\rpar^2
  &\leq 2 \left| \left(\phi, V^{-b}AV^b\phi\right) \right| \\
  &= 2 \left| \left(V^b\phi, AV^b\phi\right) \right| \\
  &\leq 2\lpar A \rpar \lpar V^b\phi \rpar^2 \\
  &= 2 \lpar A \rpar \lpar \phi \rpar^2 \\
  &< \infty
\end{split}
\end{equation*}
and thus we see that $C^{1/2}$ is $V$-smooth.

That $V$ is absolutely continuous on $R(C^{1/2})$ follows directly from
Theorem~\ref{thrm:r_ranA_ac}.\proofend


\section{\label{sec:r_finite}Finite rank perturbations}

Here, we utilise the results of \refsec{sec:r_spectra} to show that
perturbations of the form \refeq{eq:r_rankNpert} lead to the Floquet operator
having pure point spectrum for a.e.\ perturbation strength $\lambda$.

We use directly the definition of \emph{strongly $H$-finite} from Howland.


\begin{defn}[Strongly H-finite]
\label{def:r_strongHfinite}
Let $H$ be a self-adjoint operator on $\mathcal{H}$ with pure point spectrum,
$\phi_n$ a complete orthonormal set of eigenvectors, and
$H\phi_n = \alpha_n\phi_n$. A bounded operator
$A: \mathcal{H}\to \mathcal{K}$ is strongly $H$-finite if and only if
\begin{equation}
\label{eq:r_strongHfinite}
\sum_{n=1}^\infty |A\phi_n| < \infty\text{.}
\end{equation}
\end{defn}

If $H$ is thought of as a diagonal matrix on $l_2$, i.e.,\ $H = \sum_n
\alpha_n\left|\phi_n\rangle\langle\phi_n\right|$, and $A$ as an infinite
matrix $\{a_{ij}\}$, i.e.,\
$A = \sum_{m,n} a_{mn}\left|\phi_m\rangle\langle\phi_n\right|$, then
\refeq{eq:r_strongHfinite} says
\begin{equation}
\label{eq:r_A-strongHfinite}
\sum_n \left[ \sum_i \left|a_{in}\right|^2 \right]^{\frac{1}{2}} <
  \infty\text{.}
\end{equation}

For our purposes, we need to show that if $A$ is strongly $H$-finite, then
it is $U$-finite. To satisfy the assumption that $Q_\epsilon$ is trace class
in Lemma~\ref{lemma:r_5} (and hence also compact in Theorem~\ref{thrm:r_4}) we
also need to show that $A$ is trace class.


\begin{thrm}
\label{thrm:r_ufinite}
If $A$ is strongly $H$-finite, then given $U=e^{iTH/\hbar}$ for the period
$T$ in \refeq{eq:r_hamiltonian} and $H\phi_n = \alpha_n\phi_n$,
\begin{enumerate}
\item $A$ is trace class, and
  \label{thrm:r_ufinitea}	
\item $A$ is $U$-finite.
  \label{thrm:r_ufiniteb}
\end{enumerate}
\end{thrm}


\emph{Proof.} (\ref{thrm:r_ufinitea}) Simply consider
\begin{equation}
\label{eq:r_trA}
\tr(A) = \sum_l \langle \phi_l \left| A \right| \phi_l \rangle
  = \sum_l a_{ll} \leq \sum_l \left| a_{ll} \right|\text{.}
\end{equation}
For each term in the sum \refeq{eq:r_trA} we trivially have
\begin{equation*}
\left| a_{ll} \right| \leq \sqrt{\sum_i \left| a_{il} \right|^2}
\end{equation*}
and thus \refeq{eq:r_trA} is finite so $A$ is trace class.

\vspace{\baselineskip}
(\ref{thrm:r_ufiniteb}) Noting that
\begin{equation*}
U|\phi_n\rangle = e^{iTH/\hbar}|\phi_n\rangle
\end{equation*}
we calculate, by insertion of a complete set of states,
\begin{equation*}
\begin{split}
\sum_n &\langle \phi_n | G_\epsilon(\theta;U,A)|\phi_n\rangle \\
  &= \sum_n \langle \phi_n
    |A\frac{1}{\left(1-U^*e^{-i\theta_-}\right)\left(1-Ue^{i\theta_+}\right)}
    A^*|\phi_n\rangle \\
  &= \sum_m \frac{\langle \phi_m | A^*A | \phi_m \rangle}
    {\left|1-e^{-\epsilon}e^{iT\alpha_m/\hbar}e^{i\theta}\right|^2}\text{.}
\end{split}
\end{equation*}
The trace norm is then
\begin{equation*}
\tr G_\epsilon(\theta) = \sum_n \frac{| A\phi_n |^2}
  {\left|1-e^{-\epsilon}e^{iT\alpha_n/\hbar}e^{i\theta}\right|^2}\text{.}
\end{equation*}
If this is bounded for $\epsilon = 0$, then it is trivially bounded for all
$\epsilon > 0$. By \refeq{eq:r_strongHfinite} and a slightly modified version
of Theorem~3.1 in \cite{Howland87} this is finite a.e.\ for $\epsilon = 0$.
Thus the trace norm of $G_\epsilon$ exists as $\epsilon\rightarrow 0$,
which implies that the strong limit of $G_\epsilon$ exists and we conclude
that $A$ is $U$-finite.\proofend


\begin{thrm}
\label{thrm:r_sum_pert}
Let $U$ be a pure point unitary operator, and let $A_1,\ldots,A_N$ be
strongly $H$-finite. Assume that the $A_k$s commute with each other. Then
for a.e.\ $\lambda = (\lambda_1,\ldots,\lambda_N)$ in $\mathbb{R}^N$,
\begin{equation*}
V(\lambda) = e^{i\left(\sum_{k=1}^N \lambda_k A_k^*A_k\right)/\hbar}U
\end{equation*}
is pure point.
\end{thrm}


\emph{Proof.} This is a trivial modification of Theorem~4.3 in
\cite{Howland87}. Let
\begin{equation*}
\mathcal{K} = \bigoplus_{k=1}^N \bar{R}(A_k)\text{.}
\end{equation*}
The elements of $\mathcal{K}$ are represented as column vectors. Our
operator $A:\mathcal{H}\rightarrow\mathcal{K}$ is defined, for
$y\in\mathcal{H}$, by
\begin{equation*}
Ay =
  \begin{bmatrix}
  A_1y \\ \vdots \\ A_Ny
  \end{bmatrix}
  = \begin{bmatrix}
  x_1 \\ \vdots \\ x_N
  \end{bmatrix}
\end{equation*}
and therefore $A^*: \mathcal{K}\rightarrow\mathcal{H}$ is given by
\begin{equation*}
A^*x = A_1^*x_1 + \cdots + A_N^*x_N\text{.}
\end{equation*}
Accordingly, we introduce
$G_\epsilon(\theta): \mathcal{K}\rightarrow\mathcal{K}$, the matrix equivalent
of equation
\refeq{eq:r_Geps}
\begin{equation*}
\begin{split}
G_\epsilon(&\theta;U,A) \\
  &= A\bigl[1-U^*e^{-i\theta_-}\bigr]^{-1}
  \bigl[1-Ue^{i\theta_+}\bigr]^{-1}A^* \\
  &= \bigl\{A_i
    \bigl[1-U^*e^{-i\theta_-}\bigr]^{-1}\bigl[1-Ue^{i\theta_+}\bigr]^{-1}
    A_j^*\bigr\}_{1\leq i,j \leq N}\text{.}
\end{split}
\end{equation*}
The diagonal terms are finite a.e.\ because each $A_k$ is
$U$-finite by Theorem~\ref{thrm:r_ufinite}. The off diagonal terms are
of the form $X_1^*X_2$, and so the Schwartz inequality,
\begin{equation*}
\left|X_1^*X_2\right|^2 \leq \lpar X_1 \rpar^2 \lpar X_2 \rpar^2
\end{equation*}
ensures that they are finite a.e.\ too. Hence, $A$ is $U$-finite as every
term in the matrix $G_\epsilon(\theta;U,A)$ is a.e.\ finite as
$\epsilon\rightarrow 0$.

Our Hamiltonian may now be written as
\begin{equation}
\label{eq:r_hamiltonian_N}
H(\lambda) = H_0 + A^*W(\lambda)A \sum_{n=0}^\infty \delta(t-nT)
\end{equation}
and our Floquet operator as
\begin{equation*}
V(\lambda) = e^{iA^*W(\lambda)A/\hbar}U
\end{equation*}
where $W(\lambda) = \text{diag}\{\lambda_k\}$. In this form, the formalism
of \refsec{sec:r_spectra} is essentially fully regained, and we proceed
to apply Theorems~\ref{thrm:r_2}, \ref{thrm:r_4}, \ref{thrm:r_6} and
\ref{thrm:r_7}.

To establish the absolute continuity of the multiplication operator
$\mathbb{V}$ on the space $L^2(\mathbb{R}^N;\mathcal{H})$ we proceed as in
Proposition~\ref{prop:r_8}.

We write
\begin{equation*}
V(t_1,\ldots,t_N) = e^{ic\sum_{k=1}^N t_k A_k^* A_k / \hbar} U\text{,}
\end{equation*}
define
\begin{equation*}
D = -\sum_{k=1}^N \arctan(p_k/2)
\end{equation*}
where $p_k = -id/dt_k$, and compute
\begin{equation*}
C = \mathbb{V}[\mathbb{V}^*,D] = \sum_{k=1}^N C_k \geq 0\text{.}
\end{equation*}
In obtaining $C$ as a direct sum of the $C_k$, we have had to assume that
the $A_k$s commute with each other. This complication comes when considering
the term
\begin{equation*}
V(t_1,\ldots,t_N)V^*(t_1,\ldots,t_{k-1},y_k,t_{k+1},\ldots,t_N)
\end{equation*}
in the equivalent of \refeq{eq:r_C}. To obtain the required form of
$e^{ic(t_k-y_k)A^*_kA_k}$ we need the $A_k$s to commute\footnote{This
restriction is not required in
Howland's self-adjoint work because the summation over $k$ in the
Hamiltonian \refeq{eq:r_hamiltonian_N} enters directly, rather than in the
exponent of $V$.}.

Moving on, each $C_k \geq 0$ is equivalent to $C$ in
Proposition~\ref{prop:r_8}
and hence positive. Finally, we must show that $R(C^{1/2})$ is cyclic for
$\mathbb{V}$. This is no longer trivial as, for each $k$, while we have
$R(C_k^{1/2}) = R(A_k^*)$, the range of $A_k^*$ is not cyclic for $U$, hence
$\mathbb{V}$. To proceed, first note that
\begin{equation*}
R(A^*) = \bigcup_{k} R(A_k^*)\text{.}
\end{equation*}
Now, as argued in Howland, we can assume that $R(A^*)$ is cyclic for $U$.
To elaborate, define $\mathcal{M}(U, R(A^*))$ to be the smallest closed
reducing subspace of $\mathcal{H}$ containing $R(A^*)$. If $R(A^*)$ is not
cyclic for $U$, then $\mathcal{H} \ominus \mathcal{M}$ is not empty. 
However, as shown below, if $y\in\mathcal{H}\ominus\mathcal{M}$, then
$A^*WAy=0$, so in $\mathcal{H}\ominus\mathcal{M}$, $V(t) = U$ and is
therefore pure point trivially. Thus, we can ignore the space
$\mathcal{H}\ominus\mathcal{M}$, and restrict our discussion to
$\mathcal{M}$---i.e.,\ we may assume $R(A^*)$ cyclic for $U$.

The above relied upon showing that $A^*WAy=0$ for
$y\in\mathcal{H}\ominus\mathcal{M}$. We now prove this. If
$y\in\mathcal{H}\ominus\mathcal{M}$ and $y'\in\mathcal{M}$, then
\begin{equation*}
\langle y, y'\rangle = 0\text{.}
\end{equation*}
Given $y'\in\mathcal{M}$, there exists an $x\in\mathcal{K}$ such that
$y'=A^*x$, so
\begin{equation*}
\langle y, A^*x\rangle = 0\text{.}
\end{equation*}
That is
\begin{equation*}
\langle Ay, x\rangle = 0\text{.}
\end{equation*}
This is true for all $x\in\mathcal{K}$.
Suppose $y''\in\mathcal{H}$. Then $WAy''\in\mathcal{K}$ and so
\begin{equation*}
\langle Ay, WAy''\rangle=0\text{.}
\end{equation*}
That is
\begin{equation*}
\langle A^*WAy, y''\rangle=0\text{.}
\end{equation*}
As this is true for any $y''\in\mathcal{H}$, we conclude that $A^*WAy=0$ on
$\mathcal{H}\ominus\mathcal{M}$.

Thus, $R(A^*)$ (with $A$ acting on $L^2(\mathbb{R^N};\mathcal{H})$) may be
assumed cyclic for $U$, hence cyclic for $\mathbb{V}$.

We must finally show that $R(C^{1/2}) = R(A^*)$. We have
\begin{equation*}
R(A^*) = \bigcup_k R(A_k^*) = \bigcup_k R(C_k^{1/2})
\end{equation*}
and
\begin{equation*}
R(C) = \bigcup_k R(C_k)\text{.}
\end{equation*}
As $R(A^*) = R(A^*A)$, $R(C^{1/2}) = R(C)$ and we have shown that
$R(C^{1/2}) = R(A^*)$ as required.
\proofend


Finally, we wish to make the connection with our original aim---to show
that Hamiltonians of the form
\begin{equation}
\label{eq:r_rankNham}
H(t) = H_0 + \sum_{k=1}^N \lambda_k | \psi_k\rangle\langle\psi_k |
  \sum_{n=0}^\infty \delta(t-nT)
\end{equation}
have a pure point quasi-energy spectrum.


\begin{thrm}
\label{thrm:r_rankNpp}
Let $H_0$ be pure point, and define our time-dependent Hamiltonian as
in \refeq{eq:r_rankNham}. If $\psi_1,\ldots,\psi_N \in l_1(H_0)$, then
for a.e.\ $\lambda=(\lambda_1,\ldots,\lambda_N)$ in $\mathbb{R}^N$, the
Floquet operator
\begin{equation*}
V = e^{i\left(
  \sum_{k=1}^N \lambda_k |\psi_k\rangle\langle\psi_k|\right)/\hbar}U
\end{equation*}
has pure point spectrum.
\end{thrm}


\emph{Proof.} This theorem is just a special case of
Theorem~\ref{thrm:r_sum_pert} with the $A_k$s given by
$| \psi_k \rangle\langle \psi_k |$. Noting \refeq{eq:r_orth_states},
the $A_k$s clearly commute. As Howland
shows, $|\psi\rangle\langle\psi|$ is strongly $H$-finite if and only if
$\psi \in l_1(H_0)$. Thus Theorem~\ref{thrm:r_sum_pert} applies and the result
follows.\proofend


\section{\label{sec:r_discussion}Discussion of results and potential
applications}

Of fundamental importance in showing that the quasi-energy spectrum remains
pure point for a.e.\ perturbation strength $\lambda$, was the fact that
$\psi_k \in l_1(H_0)$. That is, if we write
\begin{equation*}
|\psi_k\rangle = \sum_{n=0}^\infty (a_k)_n | \phi_n\rangle
\end{equation*}
where the $| \phi_n\rangle$ are the basis states of $H_0$, then $\psi_k \in
l_1(H_0)$ if and only if
\begin{equation*}
\sum_{n=0}^\infty |(a_k)_n| < \infty\text{.}
\end{equation*}

If this requirement is dropped, and we only retain $\psi_k \in l_2(H_0)$,
then Theorem~3.1 in \cite{Howland87} fails and there is the possibility
that $V(\lambda)$ will have a non empty continuous spectrum. It was this
fact that Milek and Seba \cite{Milek90.1} took advantage of in showing that
the rank-$1$ kicked rotor could contain a singularly continuous spectral
component under certain conditions on the ratio of the kicking frequency
and the fundamental rotor frequency. They analysed two regimes of the
perturbation. One where $\psi \in l_1(H_0)$, in which case the numerical
results clearly showed pure point recurrent behaviour, and the other where
$\psi \in l_2(H_0)$, but $\psi \notin l_1(H_0)$. In the second case, the
authors further proved that the absolutely continuous part of the spectrum
was empty, and thus the system contained a singularly continuous spectral
component. The numerical results reflected this, with a diffusive type
energy growth being observed.

With the generalisation of Combescure's work here, namely our
Theorem~\ref{thrm:r_rankNpp}, it should now be possible to investigate the
full class of rank-N kicked Hamiltonians. A sufficient requirement for
recurrent behaviour has been shown to be $\psi_k \in l_1(H_0)$ and so
we must turn our attention to perturbations where this requirement is
no longer satisfied.

The challenge will be of course to find systems for which one can show
that the absolutely continuous part of the spectrum is empty. Such systems
would be candidates for classification as quantum chaotic systems.


\section*{\label{sec:r_ack}Acknowledgements}

This work was supported by the Australian Research Council.


\nocite{p_Ames}

\bibliographystyle{hunsrt}
\bibliography{bibliography}

\end{document}